\documentclass[aps,prb,reprint,superscriptaddress]{revtex4-1}

\usepackage{graphicx}
\usepackage{color}
\usepackage{amsmath}
\usepackage{bbm}
\usepackage{placeins}
\usepackage{xcolor}

\newcommand{\dir}{i} 

\begin{document}

\title{Anisotropy of the spin-orbit coupling driven by a magnetic field in InAs nanowires}
\author{Pawe{\l} W\'ojcik}
\email{pawel.wojcik@fis.agh.edu.pl}
\affiliation{AGH University of Science and Technology, Faculty of
Physics and Applied Computer Science,
Al. Mickiewicza 30, 30-059 Krakow, Poland}
\author{Andrea Bertoni}
\email{andrea.bertoni@nano.cnr.it}
\affiliation{CNR-NANO S3, Istituto Nanoscienze, Via Campi 213/a, 41125 Modena, Italy}
\author{Guido Goldoni}
\email{guido.goldoni@unimore.it}
\affiliation{Department of Physics, Informatics and Mathematics, University od Modena and Reggio Emilia, Italy}
\affiliation{CNR-NANO S3, Istituto Nanoscienze, Via Campi 213/a, 41125 Modena, Italy}

\date{\today}

\begin{abstract}
    We use the $\mathbf{k} \cdot \mathbf{p}$ theory and the envelope function approach to evaluate the Rashba spin-orbit coupling induced in a semiconductor nanowire by a magnetic field at different orientations, taking explicitely into account the prismatic symmetry of typical nano-crystals. We make the case for the strongly spin-orbit-coupled InAs semiconductor nanowires and investigate the anisotropy of the spin-orbit constant with respect to the field direction. At sufficiently high magnetic fields perpendicular to the nanowire, a 6-fold anisotropy results from the interplay between the orbital effect of field and the prismatic symmetry of the nanowire. A back-gate potential, breaking the native symmetry of the nano-crystal, couples to the magnetic field inducing a 2-fold anisotropy, with the spin-orbit coupling being maximized or minimized depending on the relative orientation of the two fields. We also investigate in-wire field configurations, which shows a trivial 2-fold symmetry when the field is rotated off the axis. However, isotropic spin-orbit coupling is restored if a sufficiently high gate potential is applied. Our calculations are shown to agree with recent experimental analysis of the vectorial character of the spin-orbit coupling for the same nanomaterial, providing a microscopic interpretation of the latter.
\end{abstract}

\maketitle

\section{Introduction}

    The spin-orbit (SO) interaction, which couples the spin of electrons with their momentum, is the functioning principle of many spintronic applications, including spin transistor,\cite{Koo2009,Wojcik2014} spin filters\cite{Wojcik2017,Ngo2010,Kohda2012} or spin-orbit qubits.\cite{Nadjperge2010,Berg2013} 
Recent investigations focus towards semiconductor nanowires (NWs) with strong SO interaction\cite{Kammhuber2017,Heedt2017,vanWeperen2015,Campos2018,Wojcik2019,Wojcik2018,Bringer2019,Zhang2006,Takase2019} as host materials for topological quantum computing based on Majorana zero energy modes.\cite{Alicea2012,Kitaev2003,Sarma2015,Oreg2010,Klausen2020} 
These exotic quasi-particles form at the ends of a NW as a result of the interplay between the SO coupling, Zeeman spin splitting and $s$-wave superconductivity induced in the NW by the proximity effect from a superconducting shell.\cite{Mourik2012,Albrecht2016,Sau2012}

    In general, a finite SO constant originates from the lack of the inversion symmetry. In semiconductors, this could either be an intrinsic feature of the crystallographic structure (Dresselhaus SO coupling\cite{Dresselhaus}) or induced by the confinement potential (Rashba SO coupling\cite{Rashba,Manchon2015}).
In zincblende NWs grown along the $[111]$ direction, the crystal inversion symmetry is preserved and the  Dresselhaus term vanishes.\cite{vanWeperen2015} 
On the other hand, for spintronic applications the Rashba term has the essential advantage of being tunable by external fields, e.g., using external gates attached to the NW.\cite{Nadjperge2012} 
In general, external fields interplay with the overall NW geometry, which is typically prismatic, and the value of the SO constant depends on the position with respect to the underlying substrate, the details of the dielectric configuration, as well as on the compositional details of the NW which determine the electronic states.\cite{Wojcik2018} 
For example, we have recently discussed the additional possibilities to engineer the SO constant in core-shell NWs with respect to homogeneous samples.\cite{Wojcik2019} 
Since the SO constant depends, in general, on the symmetry and localization of the electronic states, a magnetic field may also induce a finite SO constant due to orbital effects.
    
    Despite the number of experiments with measurements of the Rashba SO constant in semiconductor NWs,\cite{Kammhuber2017,Heedt2017,vanWeperen2015} the study of its anisotropy with respect to the magnetic field orientation is  limited. Recently, such a vectorial control was reported for InAs NWs which were suspended in order to eliminate the SO contribution originating from the substrate.\cite{Iorio2019} In Ref.~\onlinecite{Iorio2019} the authors tracked the non-trivial evolution of the weak  anti-localization (WAL) signal and determined the SO length as a function of the magnetic field intensity and direction. Interestingly, they observed that the average SO coupling is isotropic with respect to the magnetic field orientation and does not reveal any hallmark of the prismatic symmetry. When applying a transverse electric field by a gate, however, a 2-fold anisotropy appears, with the maximal SO length when $\mathbf{B}$ is perpendicular the electric field.
    
    Motivated by the availability of such experiments, we use the $8\times 8$ $\mathbf{k}\cdot \mathbf{p}$ method to analyze the dependence of the Rashba SO constant on the magnetic field intensity and orientation. The full vectorial character of the SO constant is taken into account by evaluating the SO coupling constants separately in different directions. While the magnetic field perpendicular to the NW axis is able to generate a finite SO constant which turns out to be isotropic at low intensity (below $\sim$ 1 T), for larger fields the SO constant shows a slight 6-fold symmetry with respect to the field orientation, due to the interplay between the orbital effects of the field and the prismatic symmetry of the NW. A back-gate potential couples to the magnetic field, which maximizes or minimizes the SO coupling depending on the relative orientation, leading to a 2-fold symmetry. We also investigate in-wire field configurations. The trivial 2-fold symmetry when the field is rotated in a plane which contains the axis, is almost completely removed by a gate potential. Our results are discussed in light of recent experiments reported in Ref.~\onlinecite{Iorio2019}.
    
The  paper  is  organized  as  follows. 
In  Sec.~\ref{sec:TheoreticalModel} the Rashba SO coefficients are derived from the $8\times 8$ $\mathbf{k}\cdot \mathbf{p}$ model within the  envelope function approximation, including the orbital effects which originate from the magnetic field. 
The effective  Hamiltonian  for the conduction electrons is derived  in  Sec.~\ref{sec:EffectiveHamiltonian} with details on the numerical method given in Sec.~\ref{sec:SOC}. 
Results of our calculations for homogeneous InAs NWs are reported in Sec.~\ref{sec:results}, with a discussion of recent experiments. 
Sec.~\ref{sec:summary} summaries our results.

\section{Theoretical model}
\label{sec:TheoreticalModel}

\begin{figure}[!h]
    	\begin{center}
    		\includegraphics[scale=0.38]{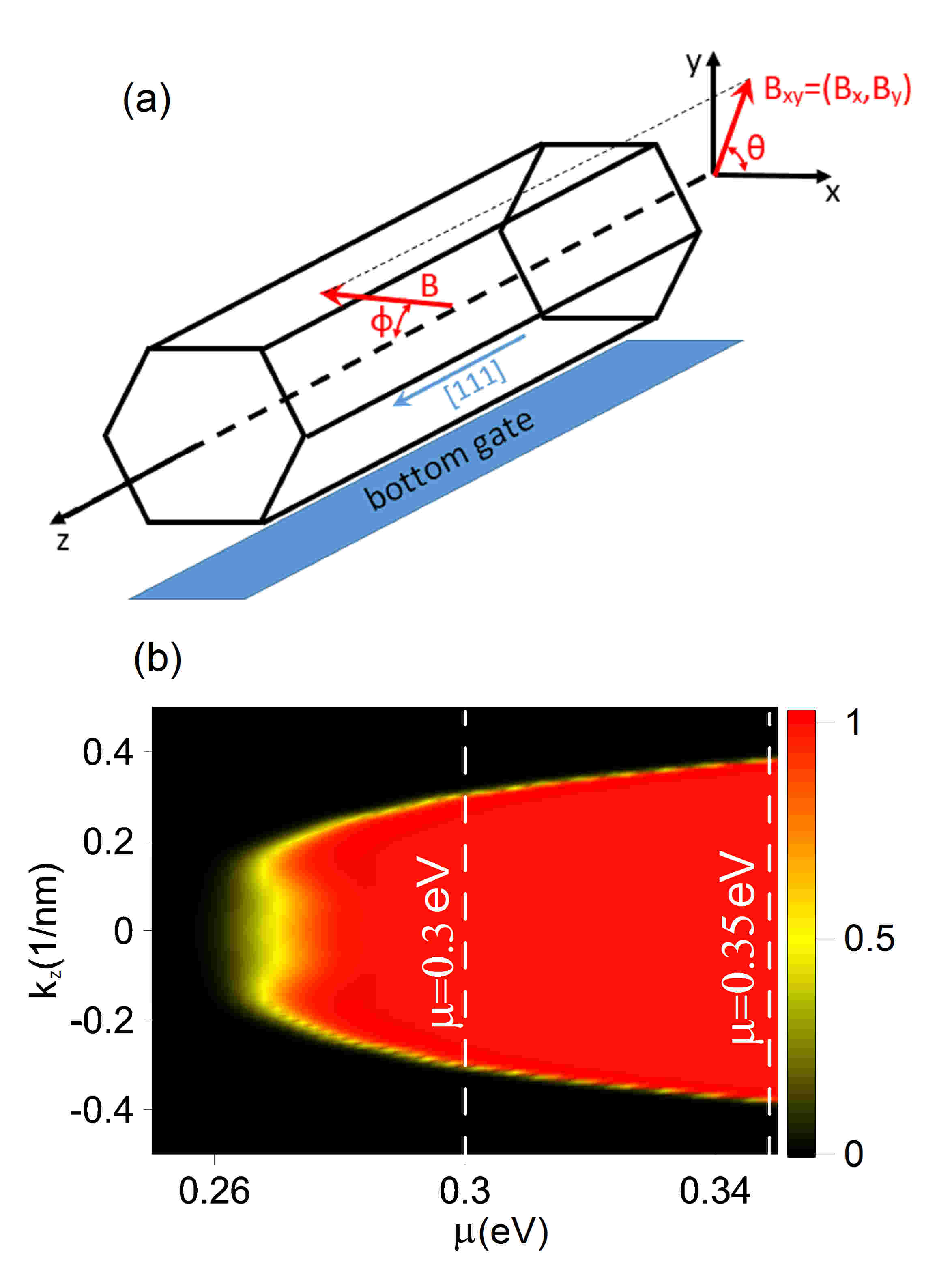}
    		\caption{(a) Schematics of a NW with a bottom gate. In our simulations, anisotropy is evaluated with a magnetic field $\mathbf{B}$ either perpendicular to the NW axis ($\phi=\pi/2$) and rotated with an azimuthal angle $\theta$, or with $\theta=\pi/2$ and rotated in the $y-z$ plane. (b) Occupation of the lowest subband as a function of the wave vector $k_z$ at each chemical potential $\mu$ at $B=4$~T. As $\mu$ increases, the occupation saturates to one at any $k_z$ below the Fermi energy. Of course, in general several subbands are occupied. The non-parabolic dispersion is clearly appreciated, with the field inducing a seemingly Landau level dispersion. Two vertical dashed lines mark values of $\mu$ selected for the further analysis.}
    		\label{fig1}
    	\end{center}
    \end{figure}

    We consider a homogeneous InAs NW with hexagonal cross-section, grown along the [111] direction for which the Dresselhaus contribution to the SO interaction can be neglected.\cite{Luo2011} The NW is subjected to the external magnetic field $\mathbf{B}=B\,(\cos \theta \sin \phi,\sin \theta \sin \phi, \cos \phi)$, with intensity $B$ and the direction being defined by the angle $\phi$ formed with the NW axis along $z$ and the angle $\theta$  formed with the $x$ axis, which connects two corners of the NW in the $x-y$ plane, see Fig.~\ref{fig1}(a). We employ the gauge $\mathbf{A}(\mathbf{r})=B\,(-y\cos \phi,0,y\cos \theta \sin \phi - x\sin \theta \sin \phi)$. A backgate is directly attached to the bottom of the NW, along a facet, generating an electric field parallel to the NW section, in the $x-y$ plane.\cite{vanWeperen2015,Wojcik2018} 
 
    Below we use the $8 \times 8$ Kane model to derive the Rashba SO constants in terms of a realistic description of the quantum states in a magnetic field. This allows for quantitative predictions of SO coefficients as a function of the magnetic field and the gate voltage for different electron 
    concentrations.\cite{Wojcik2018,Wojcik2019}

\subsection{Effective SO Hamiltonian for conduction electrons}
\label{sec:EffectiveHamiltonian}

    Our theoretical model is based on the $8 \times 8$ $\mathbf{k} \cdot \mathbf{p}$ Kane Hamiltonian within the envelope function approximation. We neglect here the spin Zeeman splitting, to focus on the dominating orbital effects, that is the distortion of the envelope function due to the field. It is straightforward to add the Zeeman splitting to the electron spin levels. The $8 \times 8$ Kane Hamiltonian reads\cite{Fabian2007} 
    \begin{equation}
    \label{eq:KH}
        H_{8\times 8}=\left ( 
        \begin{array}{cc}
        H_c & H_{cv} \\
        H^{\dagger}_{cv} & H_v
        \end{array}
    \right ),
    \end{equation}
    where $H_c$ is the Hamiltonian of conduction electrons corresponding to the $\Gamma_{6c}$ band, while $H_v$ is the Hamiltonian of the valence bands, $\Gamma _{8v}$, $\Gamma _{7v}$
    \begin{eqnarray}
    H_c&=&H_{\Gamma _6}\mathbf{1}_{2\times2},  \\
    H_v&=&H_{\Gamma _8}\mathbf{1}_{4\times4} \oplus H_{\Gamma _7}\mathbf{1}_{2\times2}.
    \end{eqnarray}
    In the above expressions
    \begin{eqnarray}
    H_{\Gamma _6}&=& -\frac{\mathbf{P} ^2}{2m_0} + E_c+ V(\mathbf{r}), \\
    H_{\Gamma _8}&=& E_c+ V(\mathbf{r})-E_0, \\
    H_{\Gamma _7}&=& E_c+ V(\mathbf{r})-E_0 - \Delta _0,
    \end{eqnarray}
    where $\mathbf{P}=\mathbf{p}-e\mathbf{A}(\mathbf{r})$, $m_0$ is the free electron mass, $E_c$ is the conduction band edge, $E_0$ is the energy gap, $\Delta_0$ is the split-off gap and $V(\mathbf{r})$ is the potential energy. In our target systems, the potential $V(\mathbf{r})$ is the sum of the Hartee potential energy generated by the electron gas and the electrical potential induced by the bottom gate attached to NW, $V(\mathbf{r})=V_H(\mathbf{r})+V_{g}(\mathbf{r})$. 

    The off-diagonal matrix $H_{cv}$ in (\ref{eq:KH}) reads
    \begin{equation}
     H_{cv}=\frac{P_0}{\hbar}\left( \begin{array}{cccccc}
             \frac{-P_+}{\sqrt{2}} & \sqrt{\frac{2}{3}}P_z & \frac{P_-}{\sqrt{6}} & 0 &
    \frac{-P_z}{\sqrt{3}} & \frac{-P_-}{\sqrt{3}} \\
    0 & \frac{-P_+}{\sqrt{6}} & \sqrt{\frac{2}{3}}P_z & \frac{P_-}{\sqrt{2}} &
    \frac{-P_+}{\sqrt{3}} & \frac{P_z}{\sqrt{3}} 
            \end{array}\right),
    \end{equation}
    where  $P_{\pm}=P_x\pm iP_y$ and $P_0=-i\hbar \langle S|\hat{p}_x|X \rangle / m_0$ is the conduction-to-valence band coupling with $|S\rangle$, $|X\rangle$ being the Bloch functions at the $\Gamma$ point of Brillouin zone.

    Finally, the folding-down transformation\cite{Fabian2007}
    \begin{equation}
    \label{eq:Hc}
         \mathcal{H}(E)=H_c+H_{cv}(H_v-E)^{-1}H_{cv}^{\dagger}.
    \end{equation}
    reduces the $8\times 8$ Hamiltonian (\ref{eq:KH}) into the $2\times 2$ effective Hamiltonian for the conduction band electrons. 

    The in-plane vector potential is introduced into the numerical model through the Peierls substitution.\cite{Nowak2018} Note that the field does not break translational invariance along the wire axis (the $z$ direction). Therefore, assuming $\Psi_{n,k_z}(x,y,z)=[\psi^{\uparrow}_{n,k_z}(x,y),\psi^{\downarrow}_{n,k_z}(x,y)]^{T}e^{ik_zz}$ and expanding the on- and off-diagonal elements of the Hamiltonian (\ref{eq:Hc}) to second order, we obtain
    \begin{eqnarray}
    \label{eq:3DH}
        \mathcal{H} &=& \bigg [ \frac{\mathbf{P}_{2D} ^2}{2m^*} + \frac{1}{2}m^*\omega_c^2 \left [ (y\cos \theta - x\sin \theta )\sin \phi - k_zl_B^2 \right 
        ]^2 \nonumber \\ &+& E_c + V(x,y) \bigg ] \mathbf{1}_{2\times 2} + (\alpha _x \sigma _x + \alpha _y \sigma _y)\frac{P_z}{\hbar},
    \end{eqnarray}
    where $\mathbf{P}_{2D}^2=P_x^2+P_y^2=(p_x+By\cos \phi)^2+p_y^2$, $\omega_c=eB/m^*$, $l_B=\sqrt{\hbar/eB}$ is the magnetic length, $\sigma_{\dir}$ are the Pauli matrices, $m^*$ is the effective mass
    \begin{equation}
         \frac{1}{m^*}=\frac{1}{m_0}+\frac{2P_0^2}{3\hbar ^2} \left ( \frac{2}{E_g} + \frac{1}{E_g+\Delta_0}
        \right ),
    \end{equation}
    and $\alpha_x$, $\alpha_y$ are the SO coefficients given by
    \begin{eqnarray}
    \label{eq:ax}
        \alpha _x(x,y) & \approx & \frac{P_0^2}{3}  \left ( \frac{1}{(E_0+\Delta_0)^2}-\frac{1}{E_0^2} \right ) \frac{\partial V(x,y)}{\partial
        y}, \\
        \label{eq:ay}
        \alpha _y(x,y) & \approx & \frac{P_0^2}{3}  \left ( \frac{1}{(E_0+\Delta_0)^2}-\frac{1}{E_0^2} \right ) \frac{\partial V(x,y)}{\partial
        x}.
    \end{eqnarray}

\subsection{SO coupling constants calculations}
\label{sec:SOC}

    Representing the Hamiltonian (\ref{eq:3DH}) in the basis of the in-plane envelope functions $\psi_{n,k_z}(x,y)$, calculated without SO coupling, i.e., the diagonal part of (\ref{eq:3DH}), the matrix elements of the SO term are given by 
    \begin{equation}
    \label{eq:a}
        \alpha_{\dir}^{nm}(k_z)=\int \int \psi_{n,k_z}(x,y) \alpha_{\dir}(x,y) \psi_{m,k_z}(x,y) dx dy.
    \end{equation}
    These coefficients define intra- ($n=m$) and inter-subband ($n\ne m$) SO constants whose magnetic field-dependence is studied in Sec.~\ref{sec:results}. Note that the $\alpha$'s coefficients depend both on the envelope functions and the gradient of the potential. 

    Calculations of the $\psi_{n,k_z}(x,y)$'s is performed by the standard self-consistent Sch{\"o}dinger-Poisson approach which includes electron-electron interaction at the mean-field level. First, the in-plane envelope functions $\psi_{n,k_z}(x,y)$ are determined from the diagonal term of (\ref{eq:3DH}) 
    \begin{eqnarray}
    \label{eq:RS2D}
         \bigg [ && \frac{\mathbf{P}_{2D} ^2}{2m^*} + \frac{1}{2}m^*\omega_c^2 \left [ (y\cos \theta - x\sin \theta )\sin \phi - k_zl_B^2 \right ]^2  \nonumber 
        \\ &+& E_c + V(x,y) \bigg ] \psi_{n,k_z}(x,y)=E_{n,k_z}\psi_{n,k_z}(x,y).
    \end{eqnarray}

    In the presence of a magnetic field, the subbands are not parabolic and $\psi_{n,k_z}(x,y)$ is explicitly $k_z$-dependent. An example of the non-parabolic dispersion is shown in Fig.~\ref{fig1}(b). Therefore, Eq.~(\ref{eq:RS2D}) is solved at selected $k_z$ on a uniform grid in $[-k_z^{max},k_z^{max}]$, with $k_z^{max}$ fairly above the Fermi wave vector. Then, the electron density is obtained by
    \begin{equation}
         n_e(x,y)=2\sum _n \int _{-k_z^{max}}^{k_z^{max}} \frac{1}{2\pi} \left| \psi_{n,k_z}(x,y) \right |^2 f(E_{n,k}-\mu,T) dk_z,
    \end{equation}
    where the factor $2$ accounts for spin degeneracy, $T$ is the temperature, $\mu$ is the chemical potential and $f(E_{n,k}-\mu,T)$ is the Fermi-Dirac distribution given by
    \begin{equation}
     f(E_{n,k}-\mu,T)=\frac{1}{1+\exp \left ( \frac{E_{n,k_z}-\mu}{k_BT} \right )}.
    \end{equation}
    Finally, for a given $n_e(x,y)$ we solve the Poisson equation 
    \begin{equation}
    \label{eq:poisson}
     \nabla ^2_{2D} V(x,y)=-\frac{n_e(x,y)}{\epsilon _0 \epsilon},
    \end{equation}
    where $\epsilon$ is the dielectric constant. 

    Equations (\ref{eq:RS2D}) and (\ref{eq:poisson}) are solved numerically on a triangular grid assuming Dirichlet boundary conditions. 
The symmetry of the discretization grid matching the symmetry of the hexagonal integration domain avoids numerical artifacts at the boundaries using smaller grid densities. 
The procedure of alternately solving Eqs.~(\ref{eq:RS2D}) and (\ref{eq:poisson}) is repeated until self-consistency is reached, which we consider to occur when the relative variation of the charge density between two consecutive iterations is lower than $0.001$ at every point of the discretization domain. 
Then, the self-consistent potential energy profile $V(x,y)$ and the corresponding envelope functions $\psi_{n,k_z}(x,y)$ are used to determine the SO constants $\alpha^{nm}_{\dir}$ from Eq.~(\ref{eq:a}).     
\begin{figure*}[!t]
    \begin{center}
        \includegraphics[scale=0.4]{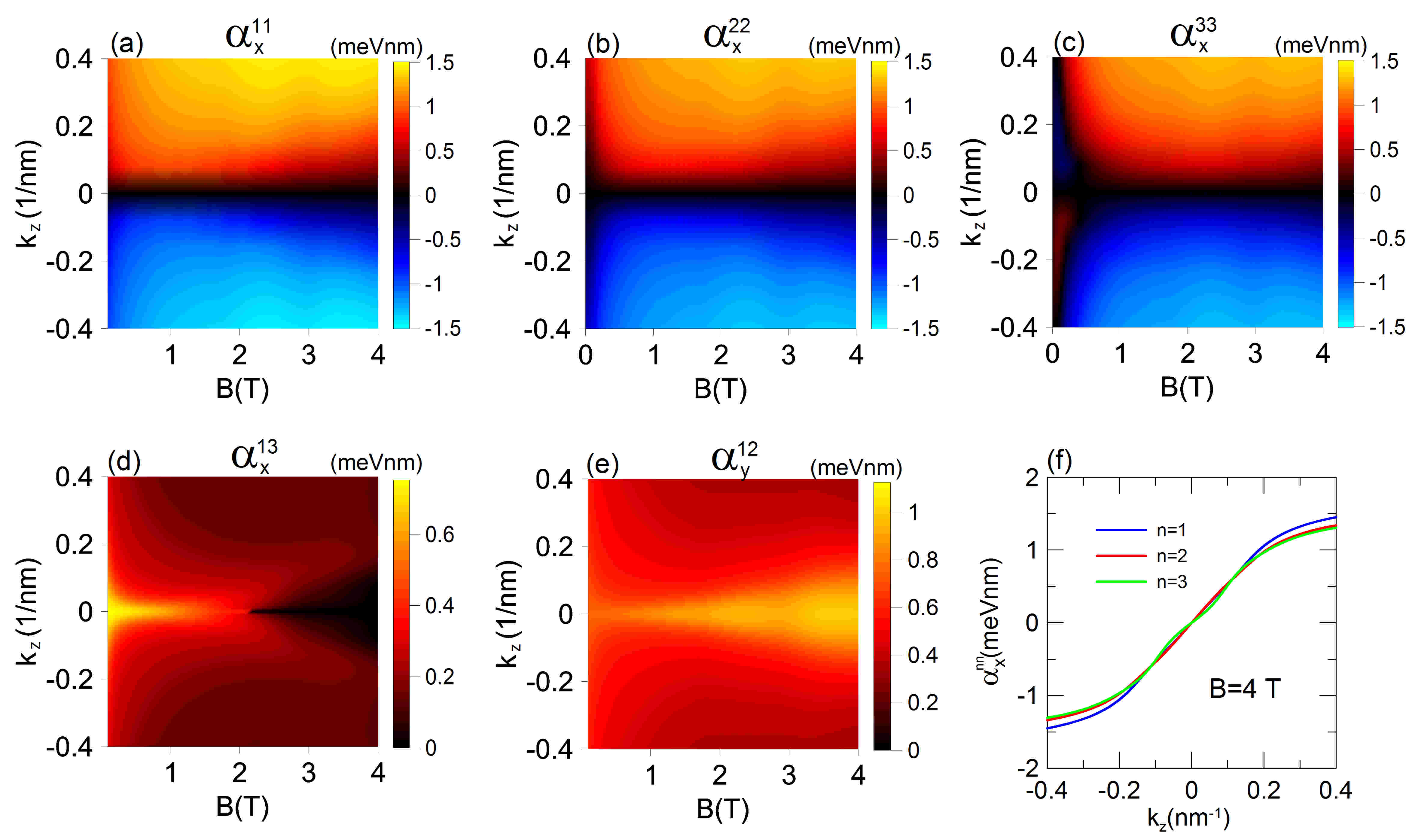}
        \caption{(a-c) Intra- ($\alpha^{nn}_{\dir}$, $n=1,2,3$) and (d,e) inter-subband ($\alpha^{1m}_{\dir}$) selected Rashba SO coupling constants as a function of the magnetic field $B$ and the wave vector $k_z$. (f) $\alpha^{nn}_{\dir}(B,k_z)$ at $B=4$~T for the three lowest states. Results are shown for $\mu=0.3$~eV and a magnetic field perpendicular to the NW axis and along the corner-corner direction ($\phi=\pi/2, \theta=0$).}
        \label{fig2}
    \end{center}
\end{figure*}
    
Further details concerning the self-consistent method for hexagonal NWs can be found in our previous papers.\cite{Bertoni2011,Royo2014}

    Calculations have been carried out for the material parameters corresponding to InAs:\cite{Vurgaftman2001} $E_0=0.42$~eV, $\Delta_0=0.38$~eV, $m^*=0.0265$, $E_P=2m_0P^2 /\hbar^2 =21.5$~eV, $\epsilon=15.15$, $T=4.2$~K, and for the NW width $W=100$~nm (facet-to-facet). In our calculations we fix the chemical potential. Results will be reported in the following section for $\mu=0.3$~eV and $\mu=0.35$~eV, which are marked by vertical dashed lines in Fig.~\ref{fig1}(b). For $B=0$, these values correspond to the electron concentration $n_e=4.8 \times 10^{16}$~cm$^{-3}$  and  $n_e=1.36 \times 10^{17}$~cm$^{-3}$, respectively. Note, however, that an increasing perpendicular magnetic field progressively depletes the NW.\cite{Royo2013} Therefore, in a transport experiment the chemical potential must be set to a sufficiently large value. In our calculations, the above two values of $\mu$ have been chosen sufficiently large as to provide an occupied ground state at the largest magnetic field intensity used here, $B=4$~T [Fig.\ref{fig1}(b)]. For a given magnetic field, different values of $\mu$ correspond to different occupations, hence a different self-consistent potential and charge distribution within the section of the NW, which in turn affects the SO coupling.

\section{Results}
\label{sec:results}

    We shall now discuss predictions of the SO constant as a function of the magnetic field intensity and direction. We shall put particular emphasis on the role of the field-induced orbital effects and the interplay with the gate potential, which also influences electronic states localization and symmetry. We conclude  this  section  by  a  discussion of the recent experiment.\cite{Iorio2019}

\subsection{Perpendicular magnetic field with no backgate potential}

    We first show that a magnetic field perpendicular to the NW axis induces  a finite Rashba SO coefficients even in the absence of any transverse electric field ($V_g=0$). In this case only the Hartree term $V_H$ contributes to the self-consistent potential. 

    For $B=0$ the self-consistent potential, having the same hexagonal symmetry of the confining potential of the NW, is symmetric with respect to the $x$ and $y$ directions. Hence, envelope functions have even or odd parity, leading to $\alpha^{nn}_x=\alpha^{nn}_y =0$ for all electronic states, as implied by Eq.~(\ref{eq:3DH}). 
    
    Let us now consider a finite magnetic field, directed along, e.g, the $x$ axis ($\phi=\pi/2,\theta=0$). The field generates an effective parabolic potential along $y$, see Eq.~(\ref{eq:3DH}), removing the symmetry of the Hamiltonian in this direction. This, in turn, induces a finite  potential gradient and a $k_z$-dependent displacement of the envelope function, hence, finite diagonal SO couplings $\alpha^{nn}_x$ [see Eq.~(\ref{eq:ax})], as shown in Fig.~\ref{fig2}~(a-c) for selected subbands. 
For a constant Fermi energy, as assumed in our calculations, the number of occupied subbands changes with magnetic field. 
At $B=1$~T, $N=8$ subbands are occupied, while only $N=3$ of them are populated at $B=4$~T. 
The behavior of $\alpha^{nn}_i(k_z)$  ($i=x,y$) for all three subbands is both qualitatively and quantitatively similar, especially for the high magnetic field, as presented in Fig.~\ref{fig2}~(f).

    The maps of $\alpha _{\dir} ^{nm}(B,k_z)$ in Fig.~\ref{fig2}~(d,e) report selected SO off-diagonal couplings between the ground state and the two lowest excited states. Other coefficients $\alpha^{1m}_{\dir}$ are four orders of magnitude lower than $\alpha _x^{11}$ and are not reported here. Note that the suppression of these off-diagonal matrix elements occurs only  for a magnetic field along the corner-corner direction, $\theta=0$. For an arbitrary  direction of the magnetic field, no symmetry applies with respect to the specific $x-y$ reference frame, and all off-diagonal SO constants have comparable values at $k_z=0$. 

    \begin{figure}[!h]
        \begin{center}
            \includegraphics[scale=0.3]{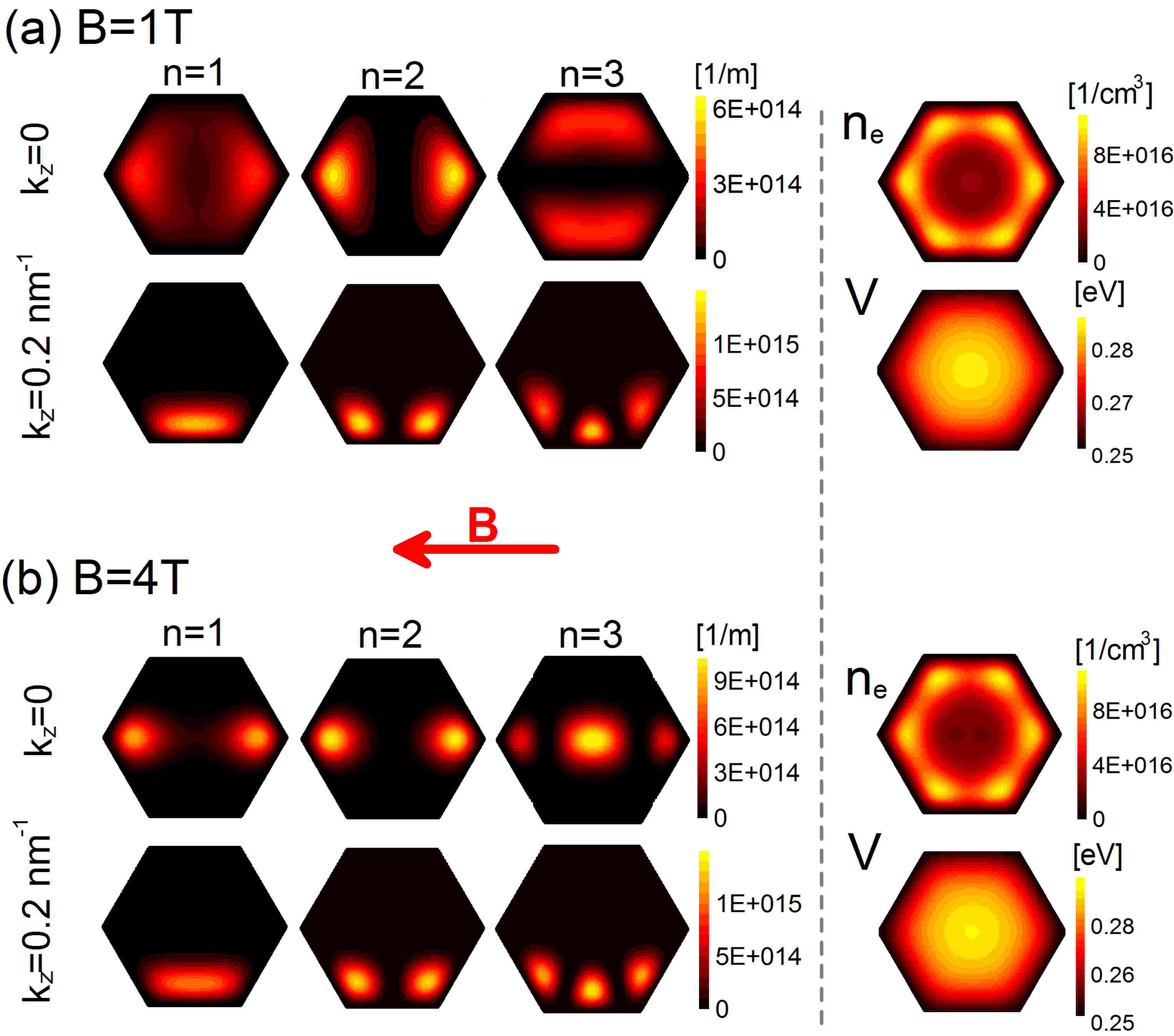}
            \caption{Squared envelope functions of the three lowest magnetic subbands, at $k_z=0$ and $0.2$~nm$^{-1}$, with a transverse magnetic field (red arrow) at intensities (a) $B=1$~T and (b) $B=4$~T. Right panels show the electron density $n_e$ and the self-consistent potential profile $V$ at the corresponding field intensities.}
            \label{fig3}
        \end{center}
    \end{figure}

    The magnetic field dependence of $\alpha^{nm}_{\dir}$ can be traced to the envelope functions localization and ensuing self-consistent potential, as shown in Fig.~\ref{fig3}. For $B=0$ (not shown) the symmetry of the envelope functions naturally leads to $\alpha^{nn}_{\dir}=0$.\cite{Wojcik2018} However, the field strongly changes the envelope function symmetry. The magnetic states of a NW have been thoroughly investigated in Ref.~\onlinecite{Royo2013}. In short, at $k_z=0$ these are localized by the field in the two corners along the field direction, where the vertical component of the field is the strongest, in seemingly dispersionless Landau levels (see also Fig.~\ref{fig1}(b)). Therefore, such states have the inversion symmetry and do not contribute to the SO coupling. At finite $k_z$ the electron states are localized at one of the facets in dispersive states, which are the analog of the traveling edge states in a Hall bar. Accordingly, the SO constant $\alpha_{x}^{nn}$ is finite, it depends on $k_z$, and changes sign at $k_z=0$, as shown in Fig.~\ref{fig2}~(a-c). Note that $\pm k_z$ states have opposite localization along $y$. Therefore, regardless of the magnetic field intensity, the self-consistent potential, which is obtained by summing states up to the Fermi wavevector, has the inversion symmetry induced by the NW confinement, as shown in the right panels of Fig.~\ref{fig3}. 

    For similar reasons, but with the opposite behavior due to symmetry, the inter-subband SO couplings  $\alpha^{1n}_{\dir}$ are largest at $k_z=0$. Its exact value strongly depends on the field intensity. Note that for the analyzed magnetic field direction the symmetry around the $y$-axis is preserved, hence $\alpha^{nn}_y=0$.

    \begin{figure}[!h]
        \begin{center}
            \includegraphics[scale=0.4]{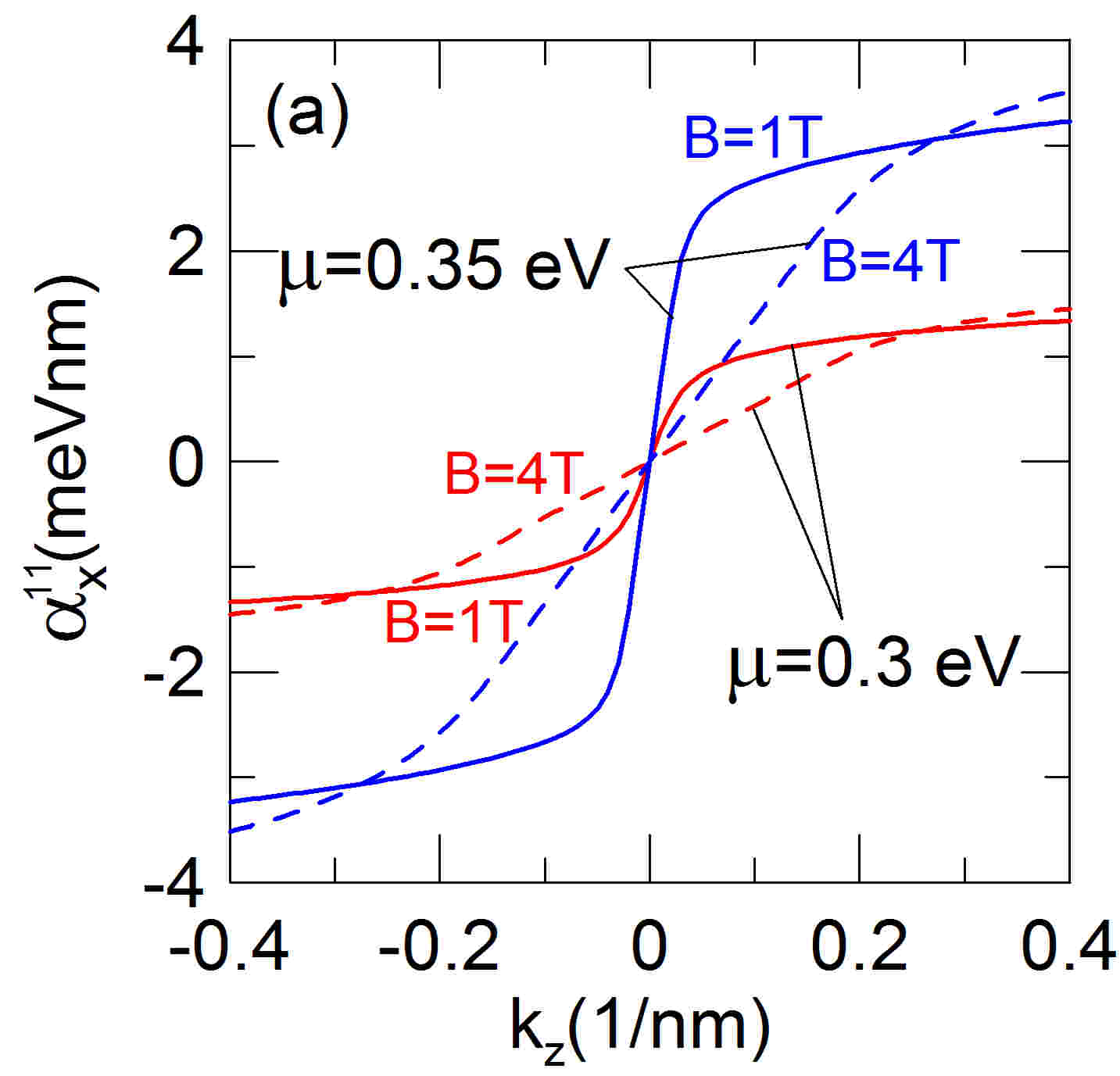}
            \vspace{0.5cm}
            \includegraphics[scale=0.4]{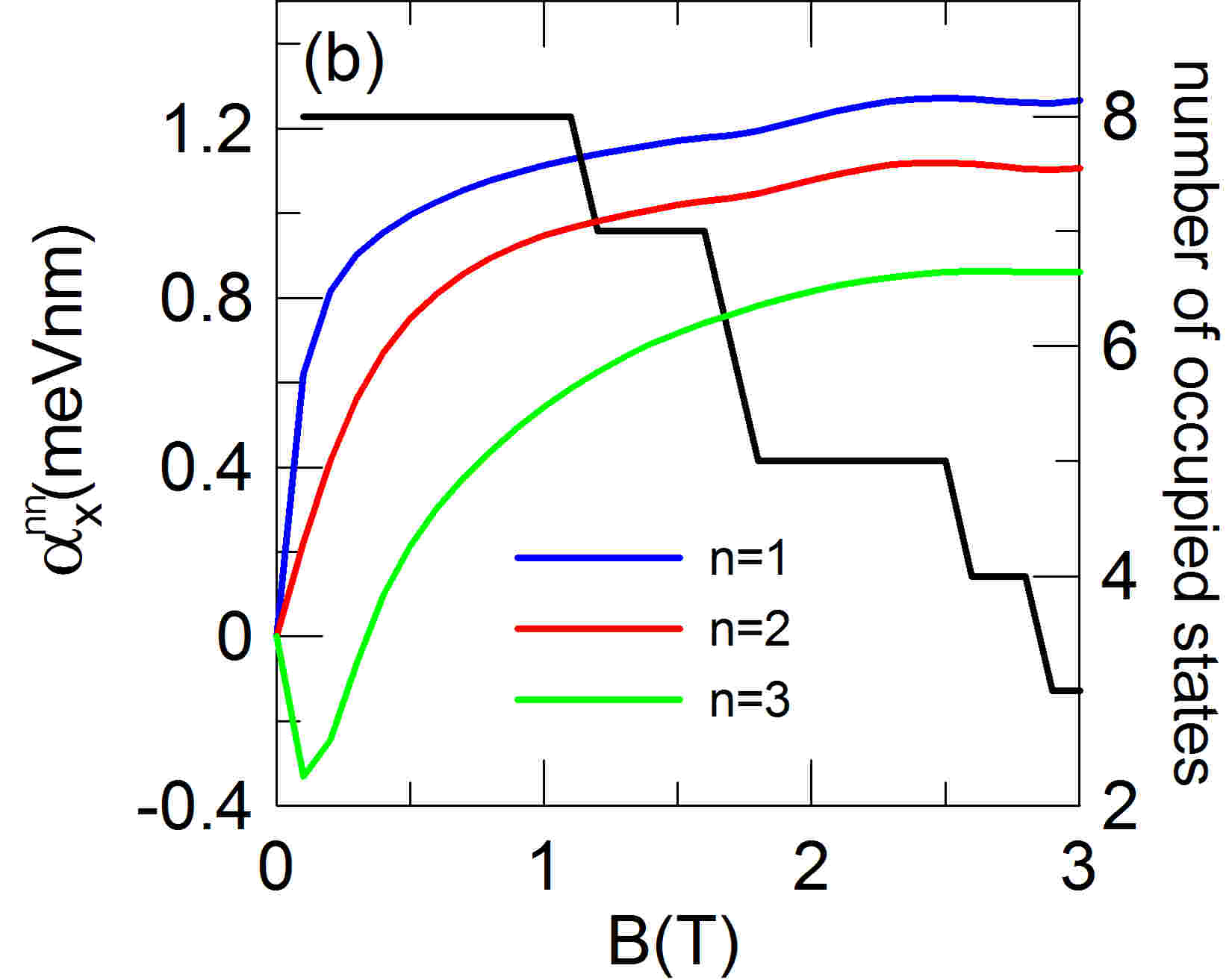}
            \caption{(a) The intra-subband SO constant $\alpha_{x}^{11}$ as a function of $k_z$ at $B=1$~T and $B=4$~T and at chemical potentials $\mu=0.30$~eV and $\mu=0.35$~eV. (b) The intra-subband SO constant $\alpha_{x}^{nn}(k_{n,z}^F), \:\:\: n=1,2,3$ (left axis) calculated at $k_{n,z}^F$ and number of occupied subbands (black line, right axis) as a function of the magnetic field intensity, $B$.}
            \label{fig4}
        \end{center}
    \end{figure}

    \begin{figure*}[!ht]
        \begin{center}
            \includegraphics[scale=0.3]{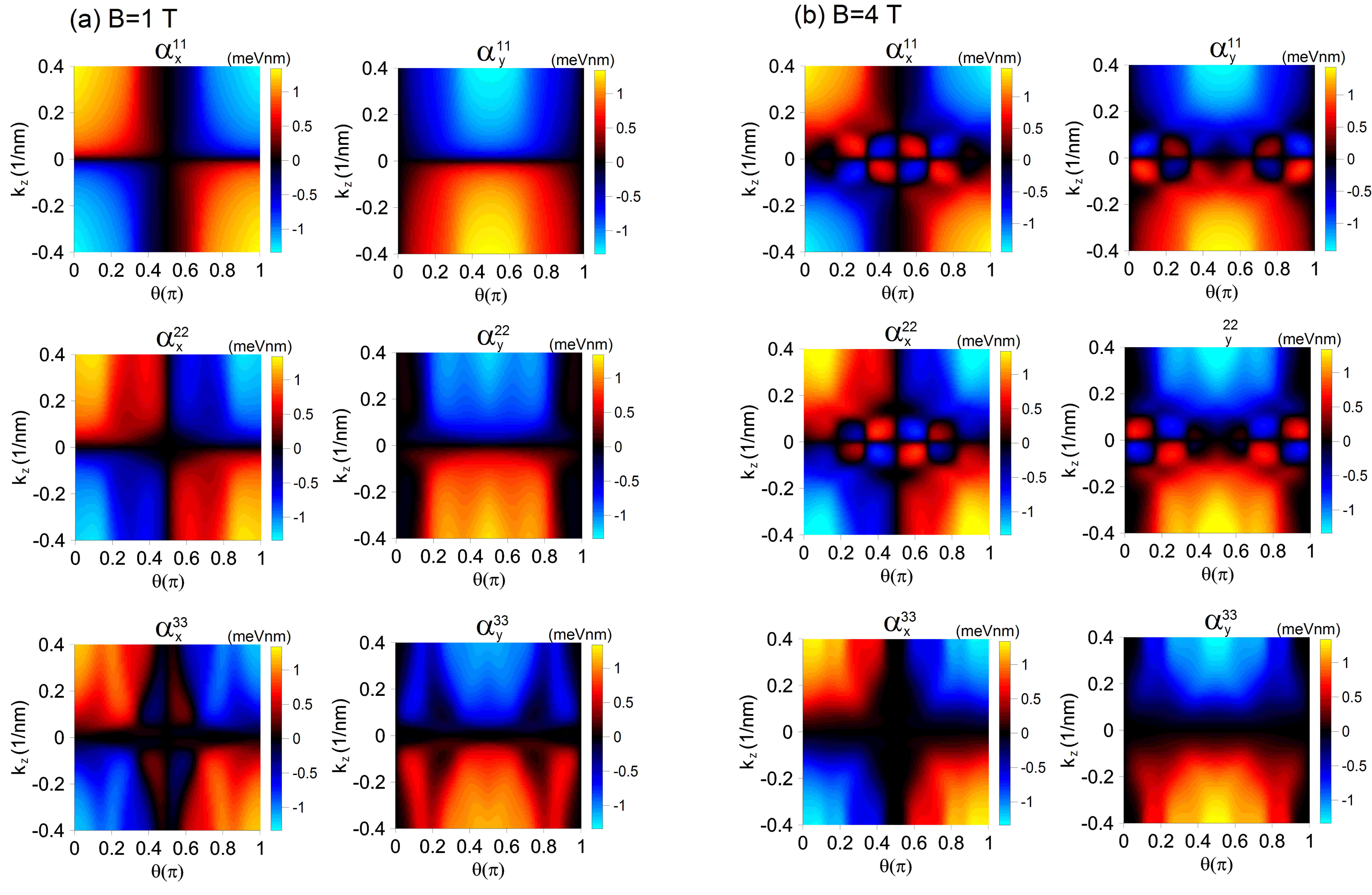}
            \caption{Maps of $\alpha^{nn}_{x}$ and $\alpha^{nn}_{y}$ as a function of $\theta$ and wave vector  $k_z$. Results are shown for $\mu=0.30$~eV and magnetic fields (a) $B=1$~T and (b) $B=4$~T.}
            \label{fig5}
        \end{center}
    \end{figure*}
    
        \begin{figure}[!ht]
        \begin{center}
            \includegraphics[scale=0.3]{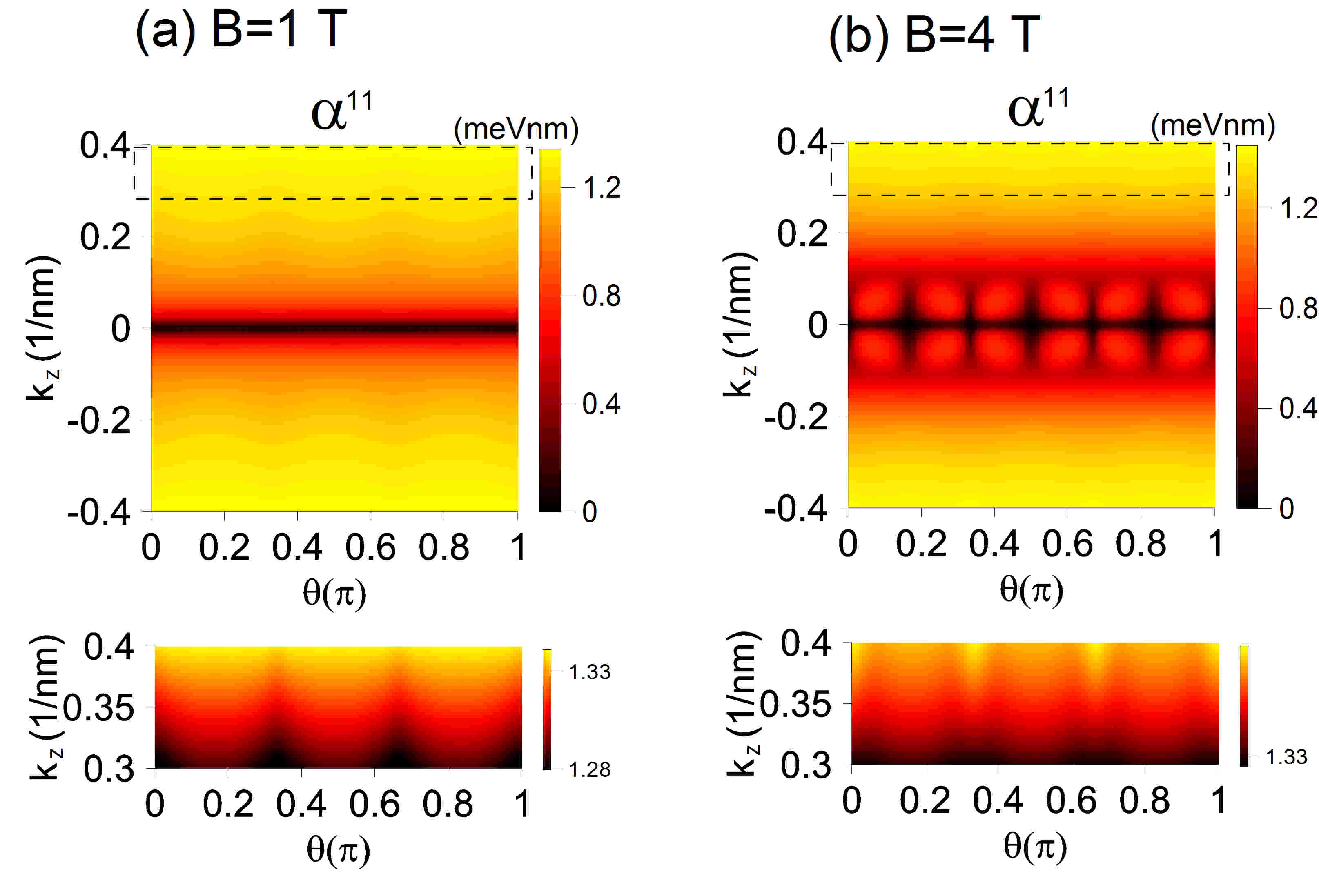}
            \caption{Maps of $\alpha^{11}$ as a function of $\theta$ and wave vector  $k_z$. Results are shown for $\mu=0.30$~eV and magnetic fields (a) $B=1$~T and (b) $B=4$~T. 
            Insets under the main panels zoom in the $k_z$ range marked by dashed black rectangle of the corresponding panel.}
            \label{fig6}
        \end{center}
    \end{figure}
    
    While a finite SO can be induced by a constant magnetic field due to the removal of the inversion symmetry, its magnitude also depends on the electric field in the NW, see Eqs.~(\ref{eq:ax}),(\ref{eq:ay}), which in turn depends on the electron concentration via the chemical potential $\mu$. At sufficiently high electron density, the free charge moves to the corners of the NW to reduce the repulsive Coulomb energy.\cite{Bertoni2011} The large gradient of the self-consistent potential where the envelope function is large generates SO constants $\alpha^{nn}_{\dir}$ which increase with $\mu$. As an example, in Fig.~\ref{fig4}(a) we show the calculated $\alpha^{11}_{\dir}$ as a function of the wavevector for $\mu=0.30$~eV and $\mu=0.35$~eV. Note that $\alpha^{11}_{\dir}$ increases rapidly with $k_z$, but then saturates as the corresponding envelope functions are squeezed more and more to the NW edges.

   In a transport experiment, electrons are injected in one of the subbands of the NW with a well defined Fermi wave vector, $k_{n,z}^F$, which is a function of the magnetic field intensity due to the field induced charge depletion. 
In Fig.~\ref{fig4}(b) we show $\alpha^{nn}_{\dir}(V_g)$ at the Fermi wave vector $k_{n,z}^F$. 
The strong localization of the electron charge at opposite NW edges gives rise to a strong susceptibility of $\alpha^{nn}_{\dir}(V_g)$ around $B=0$, analogously to what happens when a gate potential is switched on, as we discussed in Ref.~\onlinecite{Wojcik2018}. 
On the other hand, $\alpha^{nn}_{\dir}$ saturates for high magnetic fields due to the orbital effect which squeezes the envelope functions to NW edges. 
Slight oscillations of $\alpha ^{nn}_x (B)$ correspond to changes in the self-consistent potential due to depopulation of subsequent subbands when increasing field [see the black line in Fig.~\ref{fig4}(b)].

   We next analyze the anisotropy of the SO constant with respect to the transverse field direction. Indeed, as a finite $\alpha^{nn}_{\dir}$ originates from the confinement induced by the field, it is expected that the latter intertwines with the natural confinement of the electron charge at the NW edges, as discussed above. Therefore, we expect a 6-fold anisotropy with respect to $ \theta$.

    The angular dependence of the intra-subband SO couplings is shown in Fig.~\ref{fig5} for the three lowest subbands and different magnetic field intensities. Note these subbands exhaust the occupied states at $B=4$~T, but they are only a subset of the $N=8$ occupied subbands at $B=1$~T [see also Fig.~\ref{fig4}(b)]. Subbands with $N>3$ are not shown here, however, as they do not add information.

\begin{figure}[!ht]
        \begin{center}
            \includegraphics[scale=0.5]{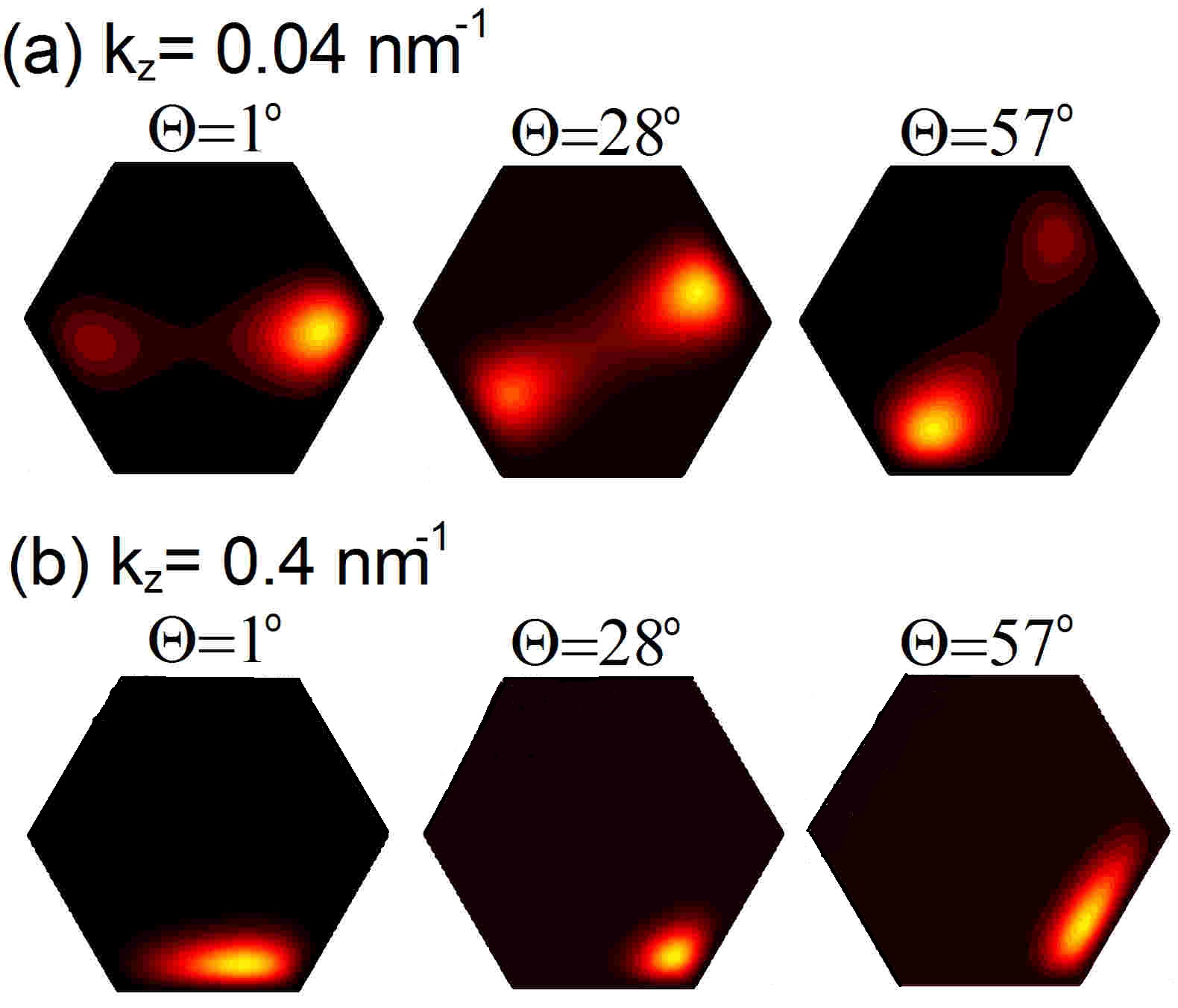}
            \caption{Squared envelope function of the ground state at selected angles $\theta$ at $B=4$~T. (a) $k_z=0.04$~nm$^{-1}$ (b) $k_z=0.4$~nm$^{-1}$.}
            \label{fig7}
        \end{center}
    \end{figure}

    In Fig.~\ref{fig6} we show $\alpha^{nn} = \sqrt{(\alpha^{nn} _x) ^2 + (\alpha^{nn} _y) ^2}$ calculated for the ground state $n=1$. The SO coupling $\alpha ^{11}$ appears isotropic and unaffected by the magnetic field orientation. However, a very weak dependence on $\theta$ can be observed in the bottom subpanels which zoom in the $k_z$ range marked by the dashed rectangular at the main graph. 
A similar weak 6-fold anisotropy is shown by all the occupied states and corresponds to the hexagonal geometry of NW. It is due to the slight reshaping of the envelope functions which localize alternately on facets and corners as the  magnetic field is rotated around the NW (see Fig.~\ref{fig7}).

    Interestingly, at $B=4$~T the SO coupling shows a flower-like pattern around $k_z=0$ for $n=1,2$, see Fig.~\ref{fig5}(b). This behavior emerges in the low $k_z$ range, where the field drives the electron charge around the NW due to the parabolic well generated by the field. However, a small $k_z$-dependent term slightly removes the symmetry, displacing the envelope function on one side and interplaying with the hexagonal potential. In Fig.~\ref{fig7}(a) the $\theta=28^\circ$ case is much more symmetric than the other two directions, due to the larger tunneling energy between the lobes, which makes the symmetric configuration more robust. In Fig.~\ref{fig7}(b), instead, the envelope function of the ground state for $k_z=0.4$~nm$^{-1}$ is strongly localized by the field near the edges. In this case the symmetry of the envelope function is strongly removed, regardless of the field direction, and only a weak anisotropy is present  thereof.

    \begin{figure}[!ht]
        \begin{center}
            \includegraphics[scale=0.2]{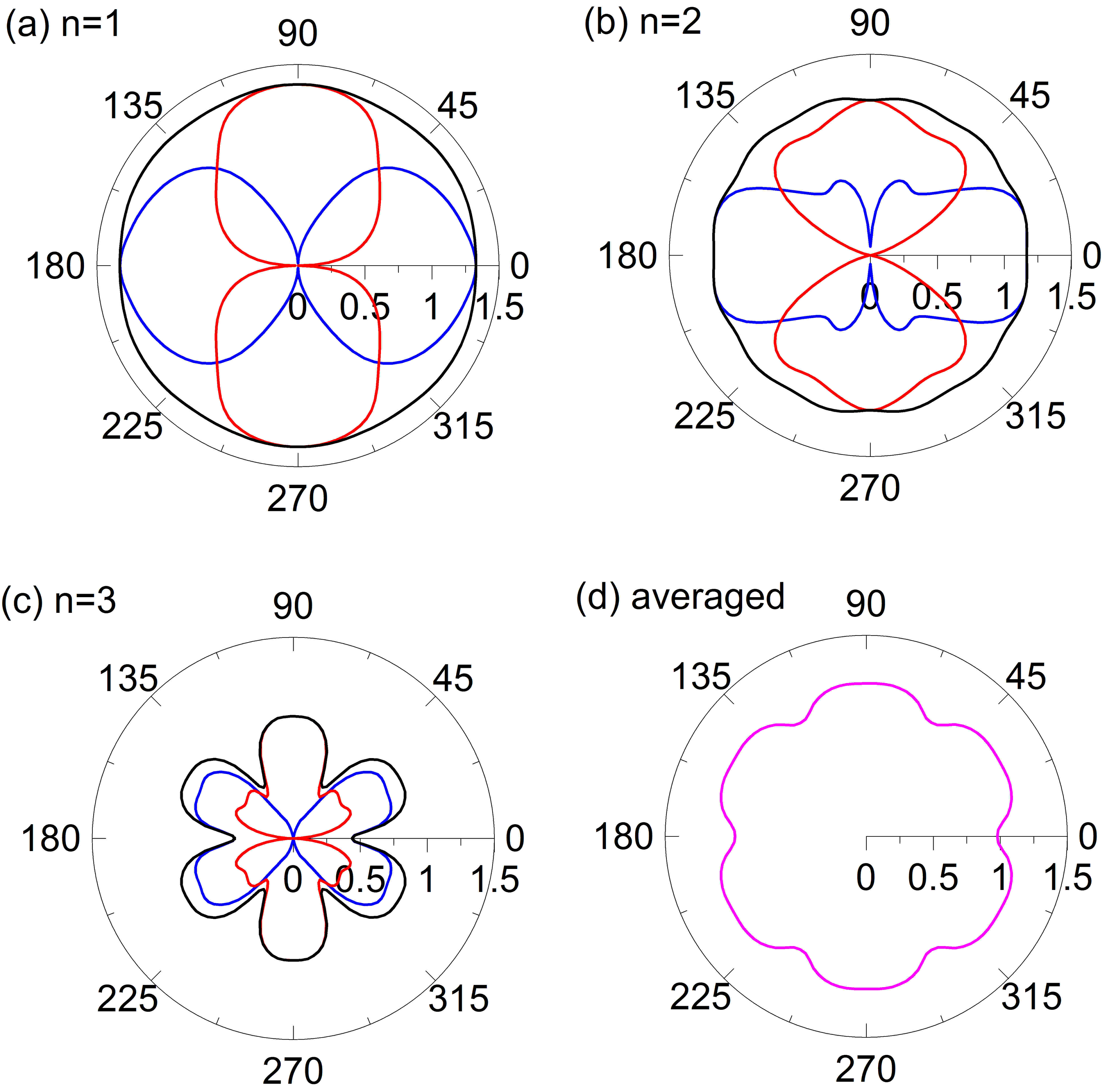}
            \caption{(a-c) Angular dependence of the $x-$ (blue) and $y-$(red) components of intra-subband SO coupling constant (in units of meVnm) at $k_{n,z}^F$ together with the modulus $\alpha ^{nn} = \sqrt{ (\alpha^{nn}_x)^2 + (\alpha^{nn}_y)^2}$ (black) for the three occupied states. Panel (d) presents the total SO coupling constant, $\alpha _{tot}$, averaged over all occupied states. Results for $B=4$~T, $\mu=0.3$~eV.}
            \label{fig8}
        \end{center}
    \end{figure}

    In Fig.~\ref{fig8}~(a-c) we report polar diagrams of the intra-subband SO constant calculated  at the Fermi wave vector $k_{n,z}^F$ for all occupied states ($N=3$) at $B=4$~T. The $x$- and $y-$ components and the modulus $\alpha^{nn}$ are shown separately. The value of SOC is the largest for the ground state, panel (a), which is almost isotropic. On the contrary, other electronic bands have a smaller values but a stronger anisotropy. The total SOC, $\alpha _{tot}$, averaged over all occupied subbands, panel (d), to be compared with the observed value in the magnetotransport experiment, shows a slight 6-fold anisotropy, with the smaller value along the corner-corner direction and the larger value along the facet-facet direction. 

    \begin{figure}[!ht]
        \begin{center}
            \includegraphics[scale=0.2]{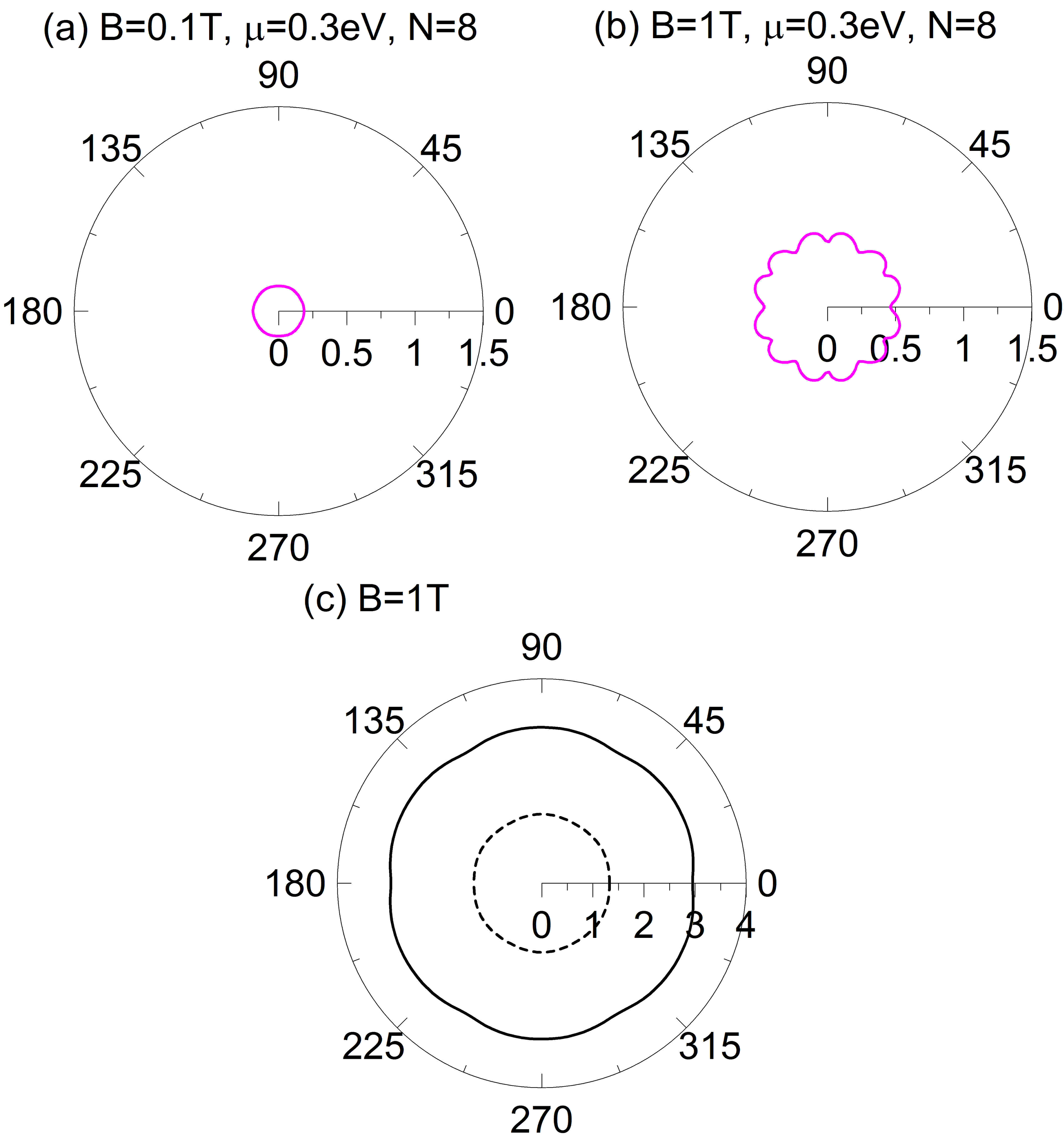}
            \caption{The angular dependence of the total SO coupling constant (in units of meVnm), $\alpha_{tot}$, averaged over all $N$ occupied subbands at $k_{n,z}^F$. (a) $B=0.1$~T, $\mu=0.3$~eV ($N=8$) and (b)  $B=1$~T, $\mu=0.3$~eV ($N=8$). (c) $\alpha^{11}$ at $\mu=0.3$~eV (dashed line) and $\mu=0.35$~eV (solid line).}
            \label{fig9}
        \end{center}
    \end{figure}
    
    The total SOC for different $B$ and $\mu$ is shown in Fig.~\ref{fig9}. 
At the lowest magnetic field $B=0.1$~T, panel (a), we do not observe any anisotropy. 
A slight 6-fold  anisotropy can be appreciated at $B=1$~T, in panel (b). 
In this case a different behaviour of the SOC as compared to that obtained at $B=4$~T is due to the averanging over a larger number of subbands ($N=8$), including higher excited states whose angular dependence is a combined effect of the orbital effects and the envelope function symmetry. 
Although the orbital effects for these higher excited states are suppressed due to low $k_{n,F}$, and therefore the contribution of them to the SOC is reduced, they cause a visible ripples of SOC, but still with the lowest SOC along the corner-corner line.  

    The observed 6-fold anisotropy of SOC is actually expected. Due to external confinement and the self-consistent field arising  from Coulomb interaction, the electron gas is strongly localized near the edges of NW for low $B$. A weak magnetic field cannot perturbate the symmetry of such strongly localized states. For higher magnetic field the Coulomb interaction weakens due to the magnetically induced charge depletion (see Fig.\ref{fig4}(b)). Therefore a sufficiently strong magnetic field may squeezee the envelope functions to the surface in a way which  depends on the relative orientation of the surface and the field. Note that the localization of the wave function at the surface is enhanced by the Coulomb repulsion at the high concentration regime. Indeed, as presented in Fig.~\ref{fig9}(c), the 6-fold anisotropy of $\alpha^{11}$ (for the ground state) is somewhat larger for higher $\mu$. 

    Our results qualitatively agree with experimental evidence in Ref.~\onlinecite{Iorio2019} where the SO coupling was measured to be isotropic in a suspended hexagonal InAs NW. 
This negative result is expected in the low magnetic field used in the experiments ($B<0.1$~T). 
Evaluating the field intensity at which anisotropy is exposed is a non trivial issue. 
The reason is that increasing the field enhances the orbital effects on the charge density, which at zero field tends to be localized near to the surface, but it also depletes the NW from free charge, which makes the charge to delocalize, due to the small Coulomb repulsion, and less sensitive to the anisotropy of the NW.

\subsection{Perpendicular magnetic field with a finite backgate potential}

    Next we consider the effect of a bottom gate attached to the NW (see Fig.~\ref{fig1}). As in the previous section, the magnetic field is perpendicular to the NW axis. We first consider the $\theta=0$ (corner-to-corner) direction, hence the two fields are orthogonal to each other.

    \begin{figure}[!htb]
        \begin{center}
            \includegraphics[scale=0.3]{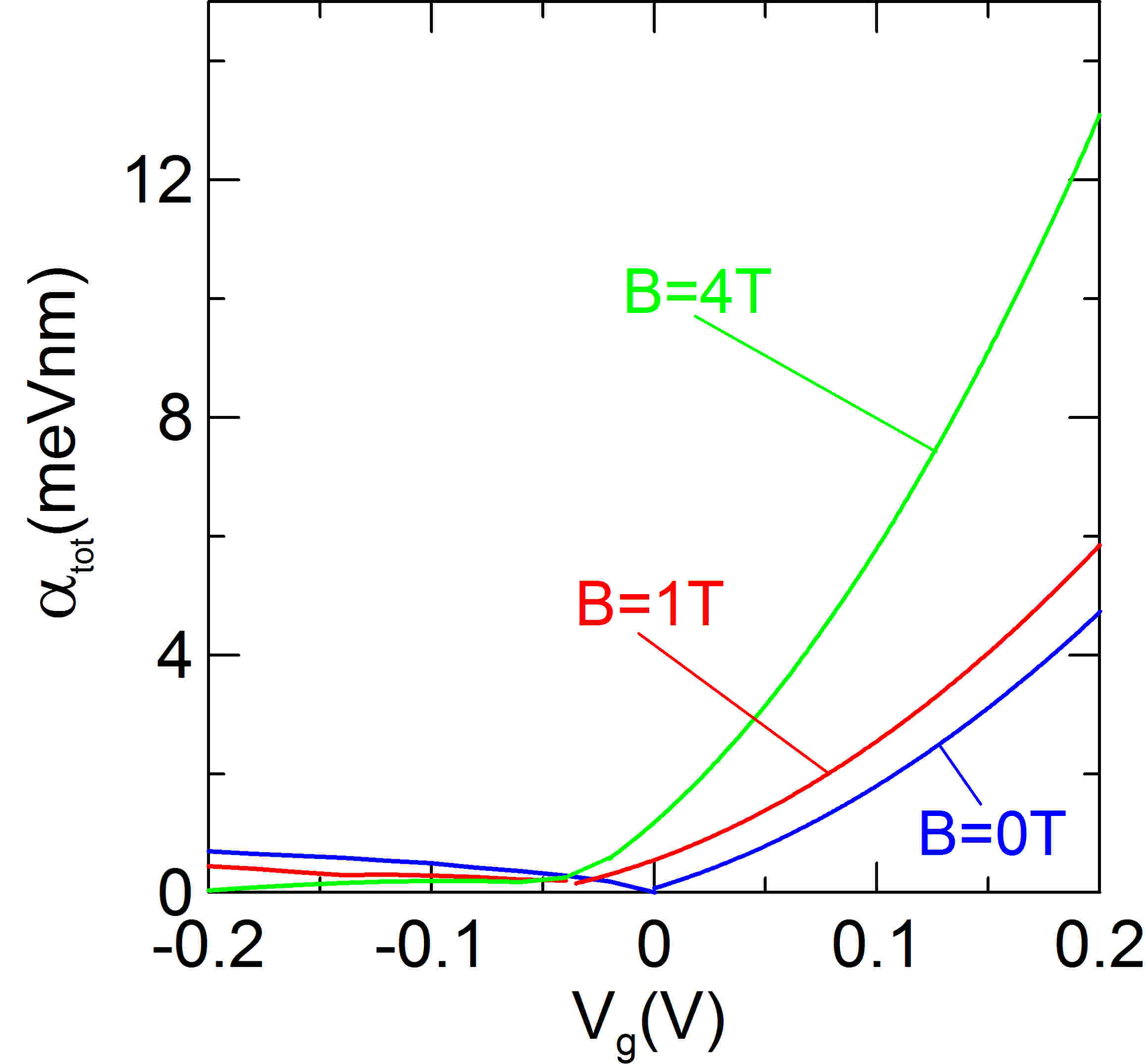}
            \caption{The total intra-subband SO, $\alpha_{tot}$, as a function of $V_g$ at selected magnetic fields $B=0, 1, 4$~T directed in the $\theta=0$ (corner-to-corner) direction. Results are shown for $\mu=0.30$~eV.}
            \label{fig10}
        \end{center}
    \end{figure}

    The total intra-subband SO coupling $\alpha _{tot}$ averaged over all occupied states at the Fermi wave vector $k_{n,z}^F$ is shown in Fig.~\ref{fig10} as a function of  the back-gate potential $V_g$ at selected field intensities. For the present fields configuration the symmetry around the $y$-axis is not broken, hence $\alpha^{nn}_y(V_g) = 0$. Figure \ref{fig10} shows that $\alpha_{tot}(V_g)$, which is finite due to the broken symmetry along $x$, increases with $B$ for $V_g>0$. $\alpha_{tot}$ takes off at a threshold $V_g$ which moves toward negative gate voltages with increasing magnetic field. 

    The strong asymmetry shown in Fig.~\ref{fig10} between positive and negative voltages is easily understood. For positive voltages the electron charge is pulled toward the gates, where the self-consistent field has the largest gradient. For negative voltages, instead, electrons are pulled far from the gate, where the potential is almost flat.\cite{Wojcik2018} Note, however, the opposite effect of the magnetic field. Here, the electric and magnetic fields are orthogonal, $\theta=0$. Therefore, for positive voltages both the gate potential and the magnetic field push electrons toward the bottom edge,  hence the magnetic field reinforces the back gate effect, increasing the SO coupling. The opposite is true for $V_g<0$; in this case, electric and magnetic field push the electrons on opposite sides, and the magnetic field weakens the SO coupling. Of course, the opposite situation takes place when the magnetic field is directed at $\theta=180 ^{\circ}$. Therefore, for a fixed $V_g$, we expect a strong anisotropy with respect to the magnetic field orientation, as shown below.
 
    \begin{figure}[!htbp]
        \begin{center}
            \includegraphics[scale=0.2]{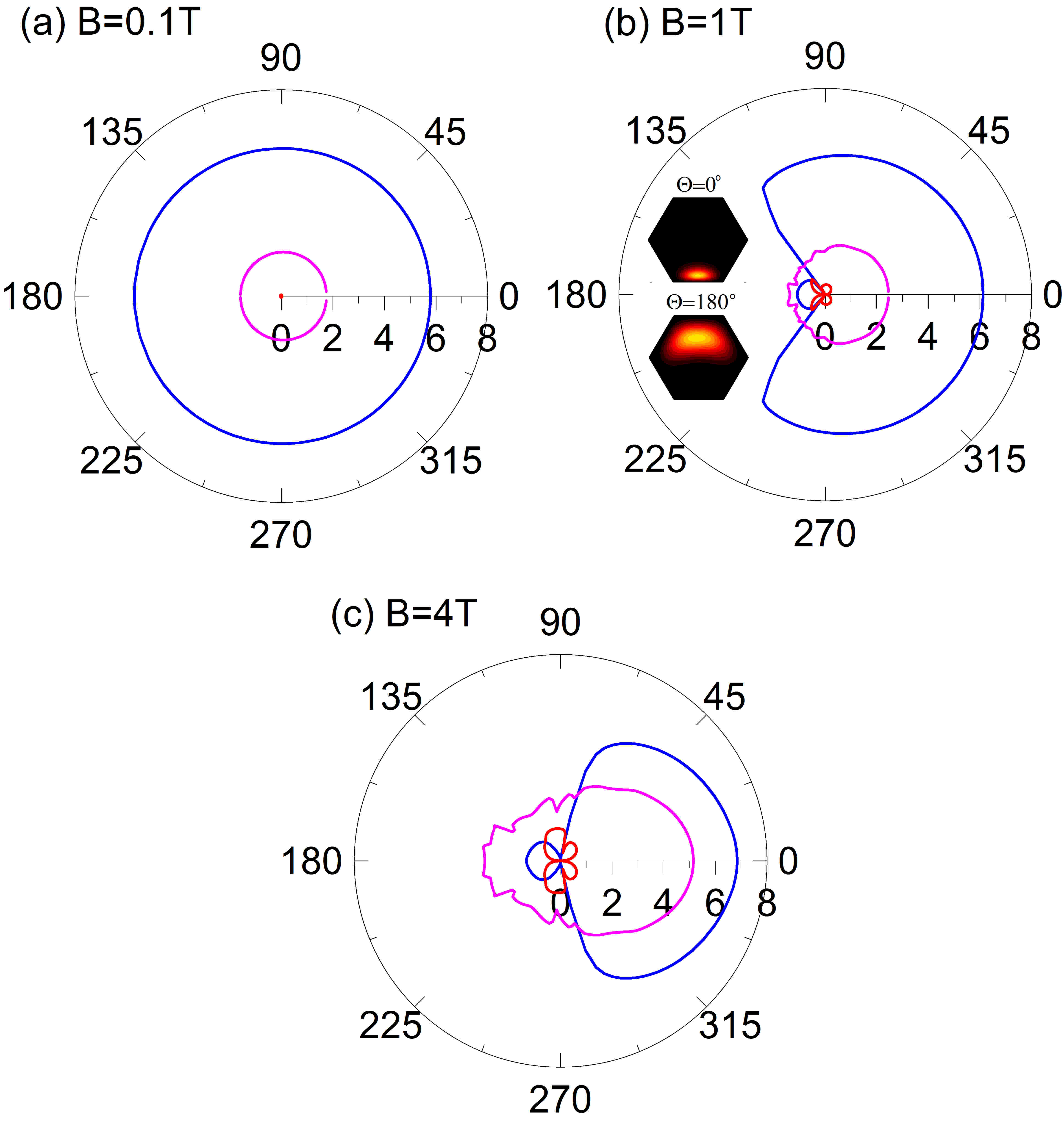}
            \caption{The angular dependence of the $x-$(blue) and $y-$ component (red) of the intra-subband SO constant (in units of meVnm) $\alpha ^{11}_{i}$ calculated at the  Fermi wave vector $k_{1,z}^F$ for the lowest subband  and the total SO constant averaged over all occupied states at $k_{n,z}^F$ (magenta line). Insets in panel (b) show the squared envelope functions of the lowest subband at $k_{1,z}^F$ for the magnetic field with $\theta=0$ and $\theta=180^\circ$. Calculations are performed with $\mu=0.30$~eV and $V_g=0.1$~eV.}
            \label{fig11}
        \end{center}
    \end{figure}

    Figure \ref{fig11} shows the polar plot of $\alpha_{tot}$ averaged over $k_{n,z}^F$ for $V_g=0.1$~V together with $\alpha_{i}^{11}$. In the absence of a magnetic field, the electronic charge is strongly localized by the electric field at the edge of the NW, near to the backgate. At a small magnetic field [$B=0.1$ T in panel  (a)], the orbital effects are negligible, and the SO coupling is isotropic. If we increase the magnetic field (panel (b)), however, $\alpha_{tot}$ (as well as $\alpha^{11}_{x}$) shows a 2-fold anisotropy, as expected from the interplay between the two fields. Note that at $\theta=180^{\circ}$, the SO coupling of the ground state is nearly zero as the orbital effects localizes the electron wave function near the upper facet (see the inset), overcoming the gate  effect.  There, the electric field is weak due to the distance from the gate, and the gradient is almost vanishing.\cite{Wojcik2018} The nonzero value of $\alpha_{tot}$ in this case results from the other states which contribute to the total SOC. Further increasing the field intensity $B$ enhances the orbital effect enhancing the anisotropy due to suppressing $\alpha^{nn}_x$ in a wide angular range, as shown in panel (c) for the ground state. 

    A similar 2-fold anisotropy has been reported in Ref.~\onlinecite{Iorio2019} with a different gate configuration, but with the same  symmetry. We postpone the detailed analysis of this experiment to Sec.~\ref{sec:ComparisonWithExperiment}. 

\subsection{Axial magnetic field}

    We now consider the SO coupling constants under a magnetic field with a component along the NW axis. This is the relevant configuration in the context of Majorana states engineering, which requires the axially magnetic field and the SO interaction to create Majorana zero energy modes at the ends of a NW. The question concerning the relative relationship between the SO coupling and the magnetic field is still an open issue.\cite{Wimmer2017}

    \begin{figure}[!ht]
        \begin{center}
            \includegraphics[scale=0.3]{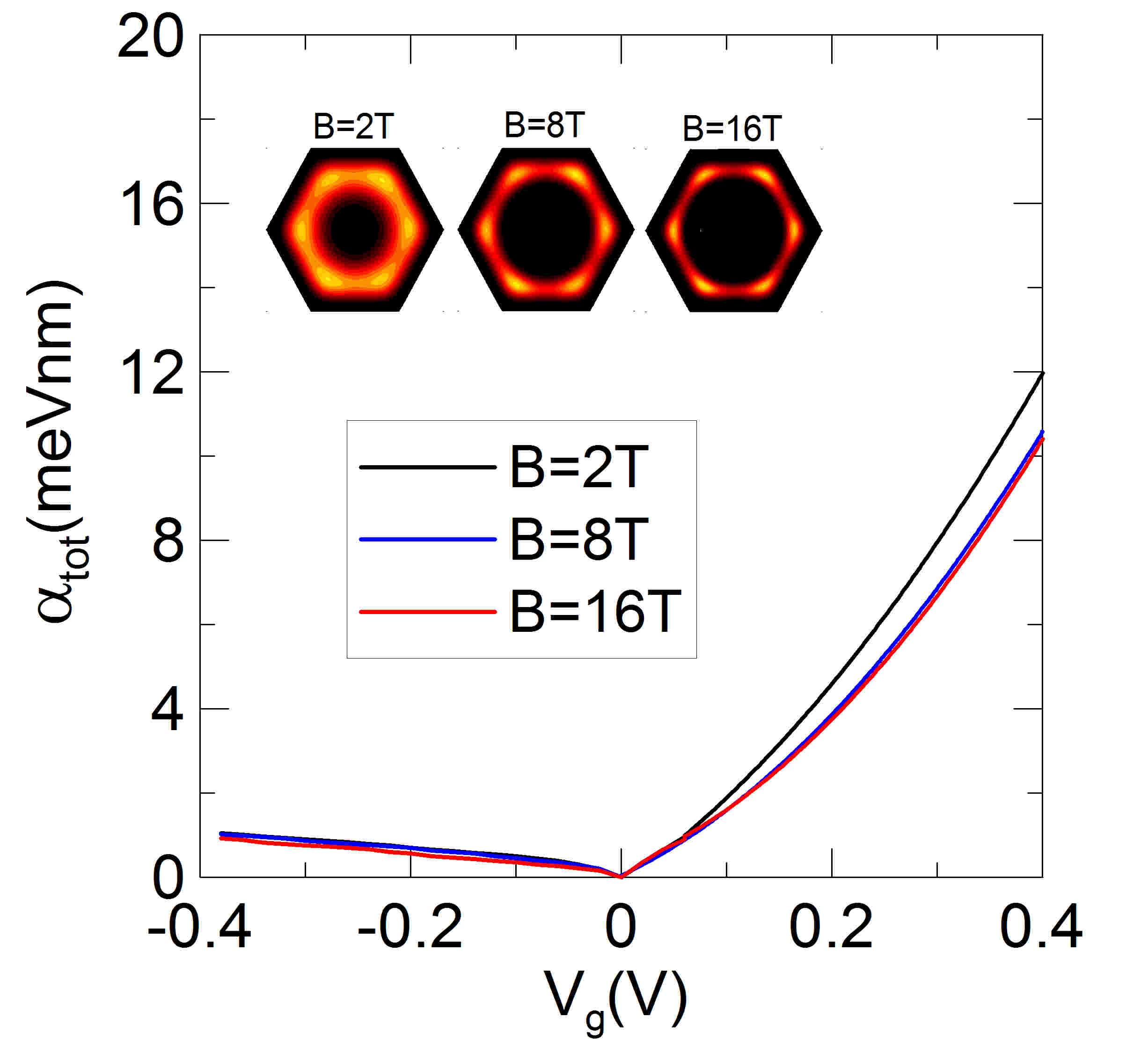}
            \caption{The total intra-subband SO constant $\alpha_{tot}$ as a function of the gate voltage $V_g$ for different axial magnetic fields. Inset: squared envelope functions of the lowest subband for different magnetic fields at $V_g=0$.}
            \label{fig12}
        \end{center}
    \end{figure}

    \begin{figure}[!ht]
        \begin{center}
            \includegraphics[scale=0.2]{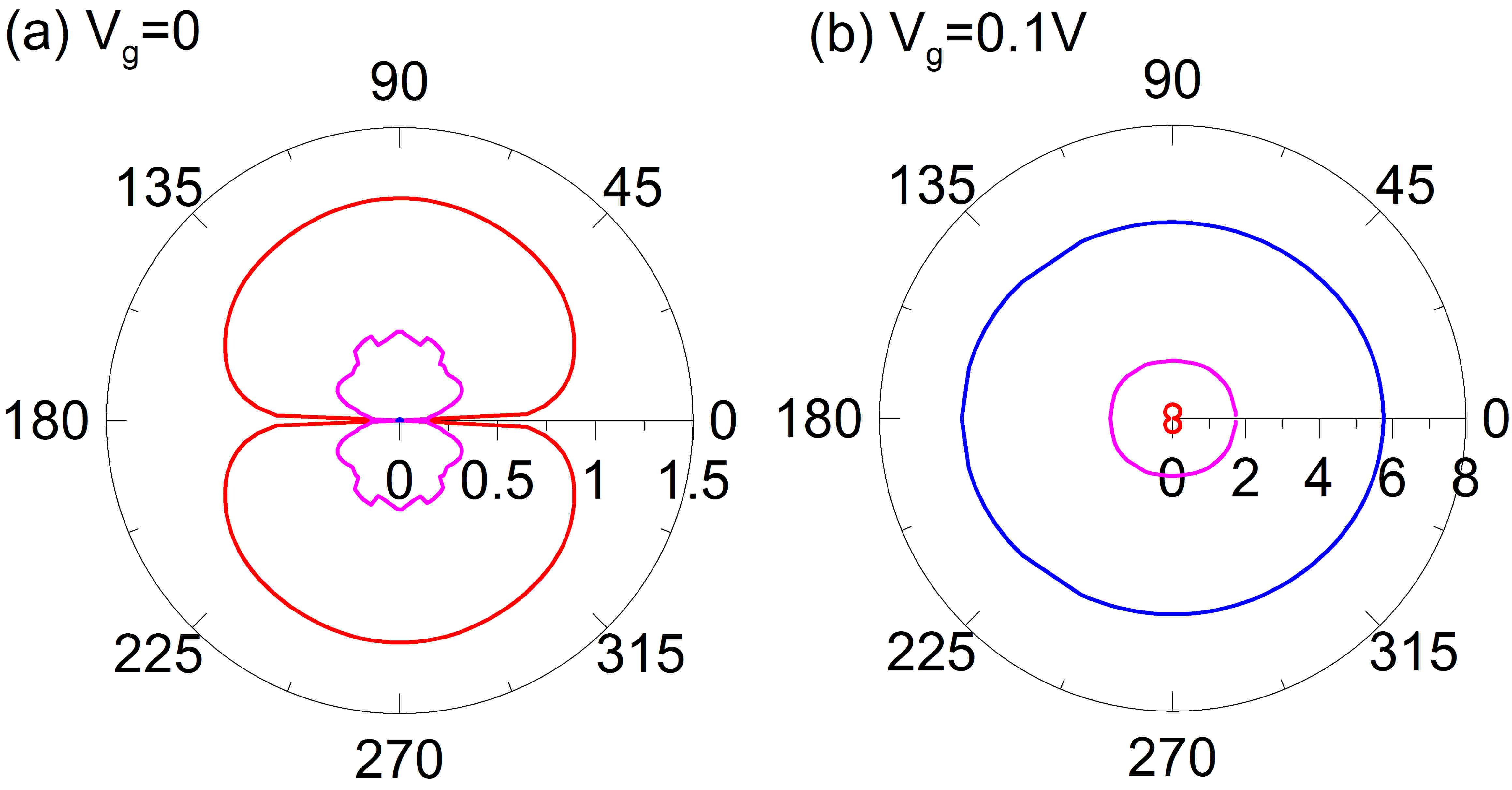}
            \caption{Angular dependence of the $x-$ (blue) and $y-$(red) component of the intra-subband SO (in units of meVnm) of the ground state $\alpha^{11}_i$ calculated at $k_{1,z}^F$ together with the total SOC $\alpha_{tot}$ (magenta). The magnetic field is rotated 
            in the $y-z$ plane. Results for $\mu=0.30$~V, $B=1$~T and (a) $V_g=0$ and (b) $V_g=0.1$~V.}
            \label{fig13}
        \end{center}
    \end{figure}

    Figure \ref{fig12} shows the calculated $\alpha_{tot}(V_g)$ \textit{vs} field intensity $B$ with an axial field ($\phi=0$). Clearly, the axial magnetic field affects the SO coupling to a slight extent up to $B=16$~T. This is in agreement with previous calculations within the Spin Density Functional formalism.\cite{Royo2015} Indeed, in the axial field configuration,  the inversion symmetry is not removed (see Eq.~\ref{eq:3DH}), although the orbital effect is still visible in the inset of Fig.~\ref{fig12}, where the envelope function is shown to localize further at the edges with the field. There is almost no field-induced depletion effect here, which is only due to the part of the orbital effect related with the field-induced quadratic terms in Eq.~\ref{eq:3DH}. Note the strong asymmetry with respect to the gate potential, which has the same explanation as the one in Fig.~\ref{fig10}.

    Next, we consider a magnetic field rotating in the $y-z$ plane, see Fig.~\ref{fig13}, which shows a 2-fold anisotropy. However, the anisotropy is almost removed by the gate potential, with the SO constant being only slightly larger for the axially magnetic field. 

    The behaviour shown in Fig.~\ref{fig13} is easily traced to the wave function localization. At $V_g=0$, SOC is trivially zero if the magnetic field is in the axial direction (inversion symmetry holds), while it is at maximum with the field in the orthogonal direction, $\phi=\pi/2$, as discussed in the previous paragraphs. If $V_g=0.1$~V, instead, the wave function is localized near to the bottom edge, where the electric field is the largest, and the SO coupling is large as well. At $B=1$ T the magnetic field does not change the localization, although if the magnetic field is perpendicular to NW the orbital effects squeezes the wave function to the side edges (either to the right or to the left) where the electric field is lower, slightly lowering the SO coupling. Hence, a small gate potential restores the $y-z$ isotropy.

\subsection{Comparison with experiment [Ref.\onlinecite{Iorio2019}]}
\label{sec:ComparisonWithExperiment}
    \begin{figure*}[!th]
        \begin{center}
            \includegraphics[scale=0.5]{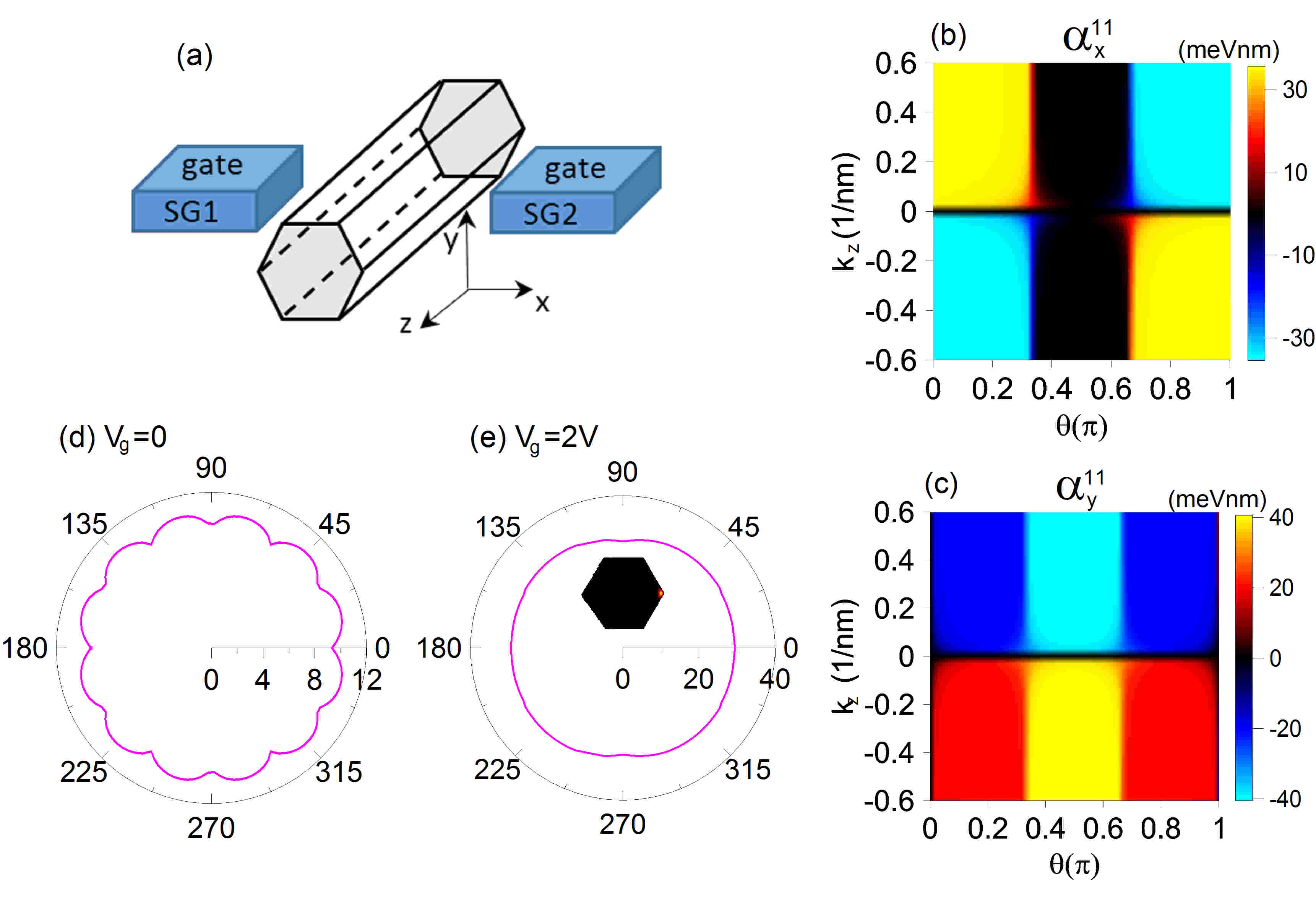}
            \caption{(a) Schematic illustration of the experimental setup. (b,c) The $x-$ and $y-$ component of the intra-subband SO coupling 
            $\alpha ^{11}_{\dir}$ as a function of the angle $\theta$ and the wave vector $k_z$. The magnetic field is rotated in the $x-y$ plane. (d,e) The angular dependence of the total SO constant, $\alpha _{tot}$. Results for $B=0.1$~T and $\phi=\pi/2$. }
            \label{fig14}
        \end{center}
    \end{figure*}
    In Ref.~\onlinecite{Iorio2019} the authors used magnetotransport experiments to determine the SO coupling in suspended InAs NWs. Using a vectorial magnet, the non-trivial evolution of weak anti-localization (WAL) is tracked and the SO length is determined as a function of the magnetic field intensity and direction. This study shows no anisotropy related to the geometrical confinement in a low field regime. The isotropy of SO coupling is however removed in the presence of an external electric field induced by side gates. In this case, the SO coupling demonstrates a 2-fold periodic angular modulation when the magnetic field is rotated in both the $y-z$ and $x-y$ plane. 

    To simulate the experimental conditions, we consider a InAs NW attached to two side electrodes located $200$~nm from the NW, see Fig.~\ref{fig14}(a). Potentials applied to the gates generate an electric field which is assumed to change linearly in the region between the electrodes. 
All parameters are taken from the experiment. 
We assume $W=100$~nm (facet-facet) and $n_e=2\times 10^{18}$~cm$^{-3}$, which for the considered NW geometry, gives $E_F=0.935$~eV. 
In order to keep the electron density constant, the field is induced by applying an asymmetric potential $V_{SG1}=\alpha_g V_{SG2}$, where $\alpha _g$ is determined separately for each $V_g$, as to keep the density constant. 
We consider only the case with the magnetic field directed perpendicular to the NW and rotating in the $x-y$ plane, with $B=0.1$~T as used in the experiment. 

    The $x$ and $y$ components of the intra-subband SO coupling for the ground state $\alpha ^{11}_i$ calculated at $V_g=0$ is presented in Fig.~\ref{fig14}(b,c). The rapid switch between the two components results from the Coulomb interaction. At the considered high electron concentrations the electron-electron repulsion localizes the charge in quasi-1D channels at the corners.\cite{Bertoni2011} When the magnetic field rotates the localization of the  ground state suddenly moves between the corners resulting in a step-like change between the $x-$ and $y-$ components which swap their intensities. 

    The total SO coupling constant averaged over all occupied states at $k_{n,F}^z$ is presented in panels (d) and (e) for two different gate voltages. The total SO coupling at $V_g=0$, panel (d), is nearly isotropic exhibiting slight oscillations with the 6-fold symmetry due the prismatic symmetry of the NW which, in the considered high electron density regime, is more pronounced due to the strong localization of electrons at the six corners. Note that in Ref.~\onlinecite{Iorio2019} the authors reported full isotropic behaviour of SOC at $V_g=0$ without the oscillations. This inconsistency remains to be clarified. It may be the result of the specific extraction of the SO length used in Ref.~\onlinecite{Iorio2019} which includes the correction from the effective NW width. Alternatively, a low resolution of the magnetotransport measurement might not be able to capture small changes of SOC. 

    Finally, we apply a potential $V_g=2$~V, as in the experiments, to the side electrodes ($\alpha_g=0.96$). In this configuration, the $y-$ component of SO coupling becomes dominant and is barely affected by the magnetic field orientation. For such a high gate potential the wave function of the ground state is strongly localized in the right corner [see the inset Fig.~\ref{fig14}(e)] and it is only slightly disturbed by the orbital effects originating from the weak magnetic field used in the experiment ($B=0.1$~T). This results in the slight 2-fold anisotropy of SOC, shown in panel (e), similarly as reported in the experiment.\cite{Iorio2019} Note however that the experimental evidence shows a 2-fold anisotropy with respect to the magnetic field orientation in the $y-z$ plane (although authors suggested its existence also in the $x-y$ magnetic field rotation) and its intensity is much stronger. 

    Although we did not perform explicit calculations in this configuration for such a high electron density, which implies a very large number of subbands ($\sim$ 100) and a correspondingly large numerical effort, results presented in Fig.~\ref{fig13} for a lower electron density and higher magnetic field agree with the experimental result and support the interpretation. Note however that at $V_g=0$ and the axially directed magnetic field, the inversion symmetry around either the $x$ and $y$ axis is not broken, which results in $\alpha_{tot}=0$ as presented in Fig.~\ref{fig13}(a). This scenario is however not supported by the experimental data which exhibit nonzero SOC even for the axially magnetic field. This strongly suggests the presence in the samples of an intrinsic electric field of an unknown origin, which is a source of SO coupling whose distortion by the weak magnetic field used in the experiment ($B=0.1$~T) is not possible, resulting in the isotropic SOC. An intrinsic electric field would explain also the absence of the SO coupling angular oscillations [as in Fig.~\ref{fig14}(a)] and the slightly lower value of SOC from the calculations, $\alpha_{tot} \approx 10$~meVnm, as compared with the corresponding experimental value $\alpha_{tot}^{exp} \approx 15$~meVnm. Interestingly, it might also explain the observed unexplained phase shift in the magnetoconductance measurement [see Fig.~3(c,d) in Ref.~\onlinecite{Iorio2019}] in terms of the relative alignment between the magnetic field and the resultant electric field (sum of the non-collinear intrinsic and extrinsic electric field) which changes depending on the applied voltage. 

\section{Summary}
\label{sec:summary}

    Based on the $\mathbf{k} \cdot \mathbf{p}$ theory within the envelope function approximation, we have analyzed the orbital effects of a magnetic field on the Rashba SO coupling in InAs homogeneous semiconductor NWs. The full vectorial character of the SO constant has been studied under the magnetic field magnitude and orientation.

    The Rashba SO interaction of conduction electrons in a NW is determined by the position and symmetry of the electron's wave function, which can be tuned by gate-induced electric fields as well as by the the orbital effects induced by a magnetic field. Specifically, when we apply the magnetic field perpendicular to NW the inversion symmetry of the envelope functions is broken and the wave functions is squeezed to the NW surface by a $k_z$-dependent effective potential. This effect results in a finite SO coupling, which is also sensitive to the geometrical confinement. As we have shown, at low magnetic field ($<1$~T for the considered NW), when orbital effects are weak, the SO coupling is isotropic with respect to the magnetic field in the NW section. Interestingly, the slight 6-fold anisotropy appears at higher magnetic fields (or high electron concentration), when the wave function is squeezed to the NW edges to a larger extent. 

    When a gate potential is applied in the direction orthogonal to the magnetic field, the two fields intertwin in a way which may enhance or suppress the SO coupling, depending on the relative direction, leading to a 2-fold anisotopy with respect to the magnetic field rotation in both the $x-y$ plane. 

    Finally, in light of our simulations, we have analyzed qualitatively recent experiments with  suspended InAs NWs\cite{Iorio2019} and good agreement with the experimental data has been found. However, we suggest that an unintended electric field is present in the sample, which would reconcile observations with our predictions.

    As a final remark, we note that in real devices a dielectric spacer often separates the gate from the NW, which reduces the SO constant. However, a spacer layer could change the cancellation effect, as it only lowers the internal electric field. Importantly, our study has shown no significant changes of the SO coupling with the axially magnetic field.

\section{Acknowledgement}

    This work was supported by the AGH UST statutory tasks No.11.11.220.01/2 within subsidy of the Ministry of Science and Higher Education in part by PL-Grid Infrastructure.

%

\end{document}